\documentclass[twocolumn,showpacs,preprintnumbers,amsmath,amssymb]{revtex4}

\usepackage{graphicx}
\usepackage{dcolumn}
\usepackage{bm}

\usepackage{amssymb}
\usepackage{amsmath}

\begin{document}


\title{Entropic cosmology in a dissipative universe}

\author{Nobuyoshi {\sc Komatsu}$^{1}$}  \altaffiliation{E-mail: komatsu@se.kanazawa-u.ac.jp} 
\author{Shigeo     {\sc Kimura}$^{2}$}

\affiliation{$^{1}$Department of Mechanical Systems Engineering, Kanazawa University, 
                          Kakuma-machi, Kanazawa, Ishikawa 920-1192, Japan \\
                $^{2}$The Institute of Nature and Environmental Technology, Kanazawa University, 
                          Kakuma-machi, Kanazawa, Ishikawa 920-1192, Japan}%
\date{\today}

\begin{abstract}
The bulk viscosity of cosmological fluid and the creation of cold dark matter both result in the generation of irreversible entropy (related to dissipative processes) in a homogeneous and isotropic universe. 
To consider such effects, the general cosmological equations are reformulated, focusing on a spatially flat matter-dominated universe.
A phenomenological entropic-force model is examined that includes constant terms as a function of the dissipation rate ranging from $\tilde{\mu} =0$, corresponding to a nondissipative $\Lambda$CDM (lambda cold dark matter) model, to $\tilde{\mu} =1$, corresponding to a fully-dissipative CCDM (creation of cold dark matter) model. 
A time evolution equation is derived for the matter density contrast, in order to characterize density perturbations in the present entropic-force model.
It is found that the dissipation rate affects the density perturbations even if the background evolution of the late universe is equivalent to that of a fine-tuned pure $\Lambda$CDM model.
With increasing dissipation rate $\tilde{\mu}$, the calculated growth rate for the clustering gradually deviates from observations, especially at low redshifts.
However, the growth rate for low $\tilde{\mu}$ (less than 0.1) is found to agree well with measurements.
A low-dissipation model predicts a smaller growth rate than does the pure $\Lambda$CDM model (for which $\tilde{\mu} =0$).
More detailed data are needed to distinguish the low-dissipation model from the pure $\Lambda$CDM one.

\end{abstract}

\maketitle

\section{Introduction}

Various cosmological models \cite{Weinberg1,Roy1,Bamba1_Miao1} have been suggested to explain the accelerated expansion of the late universe \cite{PERL1998ab,Riess1998_2007,WMAP2011,Planck2013}.
For example, a cosmological constant $\Lambda$ can be added to the Friedmann and acceleration equations in the $\Lambda$CDM (lambda cold dark matter) model.
Although the $\Lambda$CDM model is widely accepted, it suffers from several theoretical difficulties, such as the cosmological constant problem \cite{Weinberg1989}.
To resolve them, a time-varying $\Lambda(t)$ cosmology has been proposed, called the $\Lambda (t)$CDM model \cite{Freese1-Fritzsch1_Overduin1_Sola_2002-2004,Sola_2009,Sola_2011a,Sola_2011b,Sola_2013a,Sola_2013b,Sola_2013c,Sola_2014b,Lima_2014a,Sola_2014c}.     
An extra constant term is obtained from the integration of the renormalization group equation for the vacuum energy density \cite{Sola_2011b,Sola_2013b}. 
As a result, the standard $\Lambda$CDM model is actually one kind of $\Lambda (t)$CDM model.

A similar constant term appears in the CCDM (creation of cold dark matter) models \cite{Lima_1992e,Lima-Others1996-2008,Lima2010,Lima2010b,Lima2011,Lima2012,Ramos_2014,Jesus2014,Lima_2014b}, which assume that irreversible entropy is generated from gravitationally induced particle creation \cite{Prigogine1989}.
In addition, a constant term can also appear in viscous models \cite{Weinberg0,Murphy1,Barrow11-Barrow12,Davies3,Lima101,Zimdahl1-Brevik2,Nojiri1-Colistete1,Barrow21,Meng2-Avelino1_Odintsov4},   
if the bulk viscosity of the cosmological fluid is inversely proportional to the Hubble parameter \cite{C_Bulk}.   
In fact, both the creation of cold dark matter and the bulk viscosity can generate entropy in a homogeneous and isotropic universe \cite{C_Bulk_CCDM}.
Consequently, the Friedmann equation does not include an extra driving term, whereas the acceleration equation includes one due to dissipation, namely $f(t) = 0$ and $g(t) > 0$. 
Here $f(t)$ and $g(t)$ are drivers for the Friedmann and acceleration equations, respectively.

In $\Lambda (t)$CDM models, extra driving terms are given by $f(t)=g(t)= \Lambda(t)/3$, without having to assume any dissipation.
Interestingly, several forms of entropy have been proposed for the $\Lambda (t)$CDM model.  
For example, an entropic-force model has been examined from several viewpoints \cite{ Easson1,Easson2,Koivisto1_Cai1_Cai2_Qiu1_Casadio1-Costa1,Basilakos1,Lepe1,Koma4,Koma5,Koma6,Sola_2014a}. 
In that model, an entropic-force term (corresponding to an extra driver) can be derived from the (usually neglected) surface terms on the horizon of the universe \cite{Easson1}, without introducing new fields or dark energies. 
Instead of dark energy, the entropic-force model assumes that the horizon of the universe has a definite entropy and temperature \cite{Easson1}.
The most common entropic-force model is considered to be a particular case of $\Lambda (t)$CDM models \cite{Basilakos1,Sola_2014a}, so that the assumed entropy is reversible, such as the entropy related to the reversible exchange of energy \cite{Prigogine_1998}.   
On the other hand, other workers have proposed an entropic-force model similar to the CCDM and bulk viscous models, i.e., with $f(t) = 0$, as if irreversible entropy is assumed \cite{Koma5,Koma6}.

Entropic-force models can therefore be categorized into two main types \cite{Koma6}: $f(t) = g(t)$ and $f(t) = 0$.
These two types correspond to $\tilde{\mu} = 0$ (nondissipative) and $\tilde{\mu} = 1$ (fully dissipative), respectively, and have been systematically investigated \cite{Koma6}.
However, a more general dissipative universe (i.e., $0 \leq \tilde{\mu} \leq 1$) has not yet been analyzed.
Accordingly, the present paper extends the entropic-force model in order to closely examine a dissipative universe.
Specifically, a phenomenological model is considered that includes constant entropic-force terms. 
A constant term is expected to play an important role not only in $\Lambda(t)$CDM models \cite{Sola_2009,Sola_2011b,Sola_2013a,Sola_2013b} but also in entropic-force models \cite{Basilakos1,Koma6}. 
The constant term is also related to $\Lambda$CDM and CCDM models. 
The present study bridges the gap between the $\Lambda$CDM ($\tilde{\mu} = 0$) and CCDM ($\tilde{\mu} = 1$) models.

Furthermore, density perturbations are expected to be influenced by the dissipation rate, even if the background evolution of the universe remains the same.
Therefore, density perturbations in the modified entropic-force model are examined that includes a constant term for various dissipation rates.
Note that the entropic-force discussed here is  essentially different from the idea that gravity itself is an entropic force \cite{Padma1,Verlinde1}.

The remainder of the article is organized as follows.
In Sec.\ \ref{General Friedmann equations}, the general Friedmann, acceleration, and continuity equations are briefly reviewed. 
Next, the three equations are reformulated in terms of a dissipative universe.
In Sec.\ \ref{Modified entropic-force model}, a phenomenological entropic-force model is proposed that includes a constant entropic-force term.
In Sec.\ \ref{Density perturbations}, first-order density perturbations of the modified entropic-force model are analyzed.
In Sec.\ \ref{Results}, the influence of the dissipation rates are examined.
Finally, in Sec.\ \ref{Conclusions}, the conclusions are presented.

\section{Reformulation of the Friedmann equations} 
\label{General Friedmann equations}

In this section, the general Friedmann, acceleration, and continuity equations are reviewed, in accord with our previous works \cite{Koma4,Koma5,Koma6}.
Next, the three equations are reformulated, to analyze a dissipative universe related to irreversible entropy.
For this purpose, a homogeneous, isotropic, and spatially flat universe is considered
and the scale factor $a(t)$ is examined at time $t$ in the Friedmann--Lema\^{i}tre--Robertson--Walker metric. 
The present study focuses on a matter-dominated universe so that the pressure of the cosmological fluid is $p(t) = 0$.
Consequently, the general Friedmann and acceleration equations \cite{Koma4,Koma5,Koma6} become
\begin{equation}
   H(t)^2     =  \frac{ 8\pi G }{ 3 } \rho (t) + f(t)   
\label{eq:General_FRW01_f_0}
\end{equation}
and
\begin{align}
  \frac{ \ddot{a}(t) }{ a(t) }  &=  -  \frac{ 4\pi G }{ 3 }  \rho (t)  + g(t)                                   \notag \\
                                         &=  -  \frac{ 4\pi G }{ 3 }  \rho (t)  + f(t) +  h(t)                         ,     
\label{eq:General_FRW02_g_0}
\end{align}
where the Hubble parameter $H(t)$ is 
\begin{equation}
   H(t) \equiv   \frac{ da/dt }{a(t)} =   \frac{ \dot{a}(t) } {a(t)}  .
\label{eq:Hubble}
\end{equation}
Here $G$, $c$, and $\rho(t)$ are the gravitational constant, the speed of light, and the mass density of cosmological fluid, respectively, whereas
$f(t)$ and $g(t)$ are general functions corresponding to extra driving terms, i.e., entropic-force terms, as discussed later. 
It should be noted that, in Eq.\ (\ref{eq:General_FRW02_g_0}), $g(t)$ has been replaced by $f(t) + h(t)$. 
That is, $g(t)$ is divided into two functions, i.e., $f(t)$ and $h(t)$. 
In the present study, $f(t)$ and $h(t)$ are used and assumed to be related to reversible and irreversible processes, respectively.
We explain this interpretation in the next paragraph.

Previous work \cite{Koma6} examined the $\Lambda(t)$ type for which $f(t) = g(t)$, and the BV (bulk viscous) type when $f(t) =0$, which are similar to $\Lambda(t)$CDM models and bulk viscous (or CCDM) models, respectively.
[In the present paper, $f(t) = g(t)$ is equivalent to $h(t) =0$ because $h(t) = g(t) - f(t)$.]
The BV type presumes an irreversible entropy $S_{\textrm{irr}}$ arising from dissipative processes such as the bulk viscosity or the creation of cold dark matter.
In contrast, we can interpret the $\Lambda (t)$ type as assuming a kind of reversible entropy $S_{\textrm{rev}}$, such as that related to the reversible exchange of energy \cite{Prigogine_1998}. 
Accordingly, in the BV type for which $f(t) =0$, $h(t)$ is considered to be related to $S_{\textrm{irr}}$.
On the other hand, $f(t)$ is considered to be related to $S_{\textrm{rev}}$ in the $\Lambda(t)$ type for which $f(t) = g(t)$, i.e., $h(t) = 0$.
In principle, it is possible to consider a universe that includes both forms $S_{\textrm{irr}}$ and $S_{\textrm{rev}}$.
Based on this concept, the general acceleration equation is reformulated by setting
\begin{equation}
  f(t)  \geq 0  \quad \textrm{and}  \quad   h(t) \geq 0   .
\label{eq:g(t)_GE_f(t)_00}
\end{equation}
In the reformulation, it is assumed that $f(t)$ is a \textit{constant} entropic-force term based on reversible entropy $S_{\textrm{rev}}$. 
In contrast, $h(t)$ is assumed to be related to irreversible entropy $S_{\textrm{irr}}$ (and can be time-dependent variables).
Therefore, the general function $g(t)$ can be interpreted as the sum of the contributions from the irreversible and reversible entropies. That is, $g(t)$ is given by
\begin{equation}
 [g(t)]_{S_{\textrm{rev}} + S_{\textrm{irr}} } = [f(t)]_{ S_{\textrm{rev}} } + [h(t)]_{ S_{\textrm{irr}} } .
\label{eq:g_g_f_f}
\end{equation}
Consequently, Eq.\ (\ref{eq:General_FRW02_g_0}) can be rearranged as
\begin{align}
  \frac{ \ddot{a}(t) }{ a(t) }    &=  -  \frac{ 4\pi G }{ 3 }  \rho (t)   + f(t)          + h(t)                                                               \notag \\
                                          & =  -  \frac{ 4\pi G }{ 3 } \left ( \rho (t) - \frac{ 3 h(t)} {  4\pi G }  \right ) + f(t)                        \notag \\
                                          & =  -  \frac{ 4\pi G }{ 3 } \left ( \rho (t) + \frac{3 p_{e} }{c^2}  \right ) + f(t)                                             ,
\label{eq:General_FRW02_g_f}
\end{align}
where the effective pressure $p_{e}$ is defined as
\begin{equation}
  p_{e} \equiv  - \frac{  c^{2}  h(t)} {  4\pi G }  .
\label{eq:General_pe}
\end{equation}
Because $f(t) \geq 0$, the preceding formulation differs from the BV type for which $f(t)=0$.
In addition, Eq.\ (\ref{eq:General_FRW02_g_f}) includes an effective pressure $p_{e}$, which is related to the irreversible entropy $S_{\textrm{irr}}$.
In a matter-dominated universe (when $p=0$), the effective pressure $p_{e}$ is given by $p_{e}  = p + p_{c} = p_{c} $.
Then $p_{c}$ is interpreted as a pressure derived from $S_{\textrm{irr}}$.
(In the CCDM model \cite{Lima2011},  $p_{c}$ is a creation pressure for constant specific entropy.)

We next consider the continuity equation.
As examined in Refs. \cite{Koma4,Koma5,Koma6}, the general continuity equation can be obtained from the general Friedmann and acceleration equations because only two of the three equations are independent. 
The general continuity equation \cite{Koma4,Koma5,Koma6} in a matter-dominated universe becomes
\begin{align}
       \dot{\rho} + 3  \frac{\dot{a}}{a}  \rho      &=  \frac{3}{4 \pi G} H \left(  - f(t) -  \frac{\dot{f}(t)}{2 H }  +  g(t)      \right )      \notag \\
                                                                   &=  \frac{3}{4 \pi G} H \left(  h(t) -  \frac{\dot{f}(t)}{2 H }      \right )                   .
\label{eq:drho_General_00}
\end{align}
Note that we leave $\dot{f}(t)$ in Eq.\ (\ref{eq:drho_General_00}) although $f(t)$ is assumed to be constant.
This equation can be rewritten as 
\begin{equation}
       \dot{\rho} + 3  \frac{\dot{a}}{a}  \rho   =  (\Gamma + Q) \rho   ,
\label{eq:drho_General_01}
\end{equation}
where, using $p_{e}$ from Eq.\ (\ref{eq:General_pe}),  $\Gamma$ is given by
\begin{equation} 
    \Gamma = \frac{3 H}{4 \pi G}  \frac{ h(t)}{ \rho } = - 3 H \frac{ p_{e} }{\rho c^{2}} , 
\label{eq:Gamma_0}
\end{equation}
and $Q$ is 
\begin{equation} 
        Q =  - \frac{3}{8 \pi G}  \frac{ \dot{f}(t) }{  \rho  }  .
\label{eq:Q_0}
\end{equation}
In the present study, $Q$ is zero because $f(t)$ is assumed to be constant. 
We discuss this point again later.

In the standard $\Lambda$CDM model, the right-hand side of Eq.\ (\ref{eq:drho_General_01}) is zero because $f(t) =\Lambda /3 = \textrm{constant}$ and $h(t)=0$.
However, the right-hand side of this equation is in general nonzero \cite{Koma6}.
For example, in the $\Lambda (t)$CDM models \cite{Freese1-Fritzsch1_Overduin1_Sola_2002-2004,Sola_2009,Sola_2011a,Sola_2011b,Sola_2013a,Sola_2013b,Sola_2013c,Sola_2014b,Lima_2014a,Sola_2014c}, 
the right-hand side of Eq.\ (\ref{eq:drho_General_01}) is $Q \rho$ because $h(t)=0$, i.e., $\Gamma =0$.
This $\Lambda (t)$ type can be interpreted as an energy exchange cosmology \cite{Koma4,Koma5,Koma6} in which the transfer of energy between two fluids is assumed \cite{Barrow22}, 
such as an interacting quintessence \cite{Amendola1Zimdahl01}, an interaction between dark energy and dark matter \cite{Wang0102_YWang2014}, or an interaction between holographic dark energy and dark matter \cite{Pavon_2005}.
In contrast, in the BV type, the right-hand side of Eq.\ (\ref{eq:drho_General_01}) is $\Gamma \rho$ because $Q =0$ is obtained from $f(t)=0$.

We can also interpret entropic-force models from other viewpoints. 
For example, the cosmological equations considered here behave as if they were an extended $\Lambda (t)$CDM model in a dissipative universe. 
As another example, $f(t)$ can be interpreted as an effective dark energy \cite{Sola_2014a}, as discussed in Appendix \ref{Modified continuity equation}.

\section{Modified entropic-force model with constant terms}
\label{Modified entropic-force model}

This section briefly reviews the entropic-force terms.
Using that formulation, a phenomenological model is considered that includes constant entropic-force terms.
It is called a modified entropic-force model, as discussed later.

In entropic-force models, the horizon of the universe is assumed to have an associated entropy $S$ and an approximate temperature $T$ due to the information holographically stored there \cite{Easson1,Easson2}. 
Several entropic-force terms have been examined so far in this context \cite{Easson1,Easson2,Koivisto1_Cai1_Cai2_Qiu1_Casadio1-Costa1,Basilakos1,Lepe1,Koma4,Koma5,Koma6,Sola_2014a}. 
For example, the general entropic-force terms \cite{Koma6} can be written as 
\begin{equation}
  f(t) =  \alpha_{1} H^{2} + \alpha_{2} \dot{H} + (\alpha_{3}  H_{0})  H   +  \alpha_{4}  H_{0}^{2}     
\label{eq:f(g)}
\end{equation}
and
\begin{equation}
  g(t) =  \beta_{1} H^{2} + \beta_{2} \dot{H}  + (\beta_{3} H_{0}) H   +  \beta_{4} H_{0}^{2} , 
\label{eq:g(g)}
\end{equation}
where $H_{0}$ is the Hubble parameter at the present time.
The eight coefficients, $\alpha_i$ and $\beta_i$ for $i=1$ to $4$, are dimensionless constants, while 
$H^{2}$, $H$, and the constant terms are derived from an area entropy $S_{r2}$ \cite{Easson1}, a volume entropy $S_{r3}$ \cite{Koma5}, and an entropy $S_{r4}$ proportional to $r_{H}^{4}$ \cite{Koma6}, respectively, where $r_{H} = c/H$ is the radius of the Hubble horizon. 
Higher-order terms for the quantum corrections are neglected because the inflation of the early universe is not being considered.
A phenomenological derivation of the entropic-force term has been summarized elsewhere\ \cite{Koma6}.
Also, a detailed discussion of entropic cosmology has been presented by Basilakos \textit{et al.} \cite{Basilakos1,Sola_2014a}.

Here, $S_{r2}$ and $S_{r3}$ correspond to the Bekenstein black-hole entropy \cite{Bekenstein1} and the Tsallis--Cirto black-hole entropy \cite{Tsallis2012}, respectively.
In contrast, the meaning of $S_{r4}$ is less clear.
It can be considered a form of entropy that would arise if extra dimensions existed \cite{Koma6}.
Keep in mind that the Tsallis--Cirto entropy \cite{Tsallis2012} is distinct from the pure Tsallis entropy \cite{Tsa0,Tsa1}.

The properties of the entropic-force terms can be described according to Refs.\ \cite{Koma4,Koma5,Koma6}. 
To begin with, the $H^2$ and $\dot{H}$ terms cannot describe a decelerating and accelerating universe predicted by the standard $\Lambda$CDM model \cite{Basilakos1,Sola_2013a}. 
Basilakos \textit{et al.} have shown that it is not the $H^2$ and $\dot{H}$ terms but rather an extra constant term that describes a decelerating and accelerating universe \cite{Basilakos1}.
The role of the $\dot{H}$ terms is similar to that of the $H^2$ terms \cite{Basilakos1,Sola_2013c}.
The entropic-force model including $H^{2}$ terms does not properly describe cosmological fluctuations without the inclusion of a constant term \cite{Basilakos1}. 
In the $\Lambda(t)$CDM models, it has been reported that the extra constant term can be obtained from an integral constant of the renormalization group equation for the vacuum energy density \cite{Sola_2011b,Sola_2013b}. 
A similar constant term appears in the acceleration equation in the CCDM models \cite{Lima_1992e,Lima-Others1996-2008,Lima2010,Lima2010b,Lima2011,Lima2012,Ramos_2014,Jesus2014,Lima_2014b}.
The present authors \cite{Koma5} have shown that the entropic-force model with $H$ terms can describe a decelerating and accelerating universe, as in bulk viscous models.
However, the $H$ term is difficult to reconcile with astronomical observations of structure formations \cite{Koma6}.
Recently, Basilakos and Sol\`{a} have shown that simple combinations of pure Hubble terms, i.e., $H^{2}$, $\dot{H}$, and $H$ terms, are insufficient for a complete description of the cosmological data \cite{Sola_2014a}. 
Thus the constant term plays an important role.

Consider the most important and simplest term,
namely a phenomenological model that includes \textit{constant} entropic-force terms. 
It is here called the modified entropic-force model.
The general functions are taken to be  
\begin{equation}
  f(t) =   \alpha_{4}  H_{0}^{2}     \quad  \textrm{and}   \quad     g(t) =    \beta_{4} H_{0}^{2}    . 
\label{eq:f(Cst)_g(Cst)_00}
\end{equation}
In the present study, $g(t)$ is replaced by $f(t) + h(t)$, where $h(t)$ is given by $g(t) -f(t)$, as discussed in the previous section.
Using $h(t)$, Eq.\ (\ref{eq:f(Cst)_g(Cst)_00}) can be written as
\begin{equation}
  f(t) =   \alpha_{4}  H_{0}^{2}     \quad  \textrm{and}   \quad     h(t) =    (\beta_{4} - \alpha_{4})  H_{0}^{2} = \gamma_{4}^{\textrm{irr}}  H_{0}^{2}     , 
\label{eq:f(Cst)_g(Cst)}
\end{equation}
where it is assumed that
\begin{equation}
    \alpha_{4}    \geq 0     \quad       \textrm{and}   \quad             \gamma_{4}^{\textrm{irr}} = \beta_{4} - \alpha_{4}    \geq 0   .
\label{eq:beta4_GE_alpha4_delta}
\end{equation}
The coefficient $\gamma_{4}^{\textrm{irr}} $ is a dimensionless constant, which is assumed to be related to an irreversible entropy.
In this paper, Eq.\ (\ref{eq:f(Cst)_g(Cst)}) is used for the modified entropic-force model. 
Substituting Eq.\ (\ref{eq:f(Cst)_g(Cst)}) into Eq.\ (\ref{eq:General_FRW01_f_0}), the modified Friedmann equation becomes
\begin{equation}
   H^{2}  = \frac{ 8\pi G }{ 3 } \rho    +  \alpha_{4}  H_{0}^{2}   . 
\label{eq:FRW01(g)}
\end{equation}
Likewise, substituting Eq.\ (\ref{eq:f(Cst)_g(Cst)}) into Eq.\ (\ref{eq:General_FRW02_g_0}), the modified acceleration equation can be rewritten as 
\begin{equation}
  \frac{ \ddot{a} }{ a } =  \dot{H} + H^{2}   =  -  \frac{ 4\pi G }{ 3 } \rho    +  \alpha_{4}  H_{0}^{2} +  \gamma_{4}^{\textrm{irr}}  H_{0}^{2}   . 
\label{eq:FRW02(g)}
\end{equation}
Using Eq.\ (\ref{eq:General_FRW02_g_f}), a modified acceleration equation is obtained that is equivalent to Eq.\ (\ref{eq:FRW02(g)}),
\begin{equation}
  \frac{ \ddot{a} }{ a }   
       =  -  \frac{ 4\pi G }{ 3 } \left ( \rho  + \frac{3 p_{e} }{c^2}  \right ) + \alpha_{4}  H_{0}^{2}   , 
\label{eq:General_FRW02_g_2_alpha4}
\end{equation}
where the effective pressure $p_{e}$ is 
\begin{equation}
   p_{e} =  - \frac{  c^{2} H_{0}^{2}} {  4\pi G }  \gamma_{4}^{\textrm{irr}}   .
\label{eq:pe_Q_1_alpha4}
\end{equation}
Equation (\ref{eq:pe_Q_1_alpha4}) implies a \textit{constant} effective pressure.
Assume that the $\alpha_{4}  H_{0}^{2}$ term in Eqs.\ (\ref{eq:FRW01(g)}), (\ref{eq:FRW02(g)}), and (\ref{eq:General_FRW02_g_2_alpha4}) corresponds to the entropic-force terms derived from reversible entropy in the standard entropic-force model. 
In contrast, the effective pressure $p_{e}$ is assumed to be related to the irreversible entropy.
That is, the $\gamma_{4}^{\textrm{irr}}  H_{0}^{2}$ term in Eq.\ (\ref{eq:FRW02(g)}) is assumed to be related to the irreversible entropy.
Accordingly, Eq.\ (\ref{eq:FRW02(g)}) includes the effect of both reversible and irreversible entropies.  
(The irreversible entropy considered here is not necessarily the same as the entropy on the horizon of the universe.)
The present entropic-force phenomenology thereby constitutes an extended model. 

In the modified entropic-force model, the $\alpha_{4}  H_{0}^{2}$ term (related to reversible entropy) can be interpreted as a modification of the Einstein tensor.
In contrast, the effective pressure $p_{e}$ (related to irreversible entropy) is interpreted as a modification of the energy--momentum tensor of the Einstein equation.
The cosmological equations examined here are equivalent to those of an extended $\Lambda$CDM model in a dissipative universe, as proven in Appendix \ref{Modified continuity equation}.

Next, consider the modified continuity equation in the present model.
Substituting Eq.\ (\ref{eq:f(Cst)_g(Cst)}) into Eq.\ (\ref{eq:Q_0}), we find $Q =0$ because $f(t)$ is assumed to be constant. 
Substituting $Q=0$ into Eq.\ (\ref{eq:drho_General_01}), the modified continuity equation is found to be 
\begin{equation}
       \dot{\rho} + 3  \frac{\dot{a}}{a}  \rho   =  \Gamma  \rho   ,
\label{eq:drho_General_01_alpha4}
\end{equation}
where, substituting Eqs.\ (\ref{eq:f(Cst)_g(Cst)}) and (\ref{eq:pe_Q_1_alpha4}) into Eq.\ (\ref{eq:Gamma_0}), $\Gamma$ is given by 
\begin{equation} 
    \Gamma = \frac{3H}{4 \pi G}  \frac{  \gamma_{4}^{\textrm{irr}}  H_{0}^{2} }{ \rho } = - 3 H \frac{ p_{e} }{\rho c^{2}}     .
\label{eq:Gamma_0_alpha4}
\end{equation}
Using the effective pressure, Eq.\ (\ref{eq:drho_General_01_alpha4}) can be rewritten as 
\begin{equation}
       \dot{\rho} + 3  \frac{\dot{a}}{a} \left (  \rho + \frac{p_{e}}{c^2}  \right )   = 0   .
\label{eq:drho_General_01_pe}
\end{equation}
The above formulation is similar to the BV type \cite{Koma6}. 
However, the modified continuity equation considered here is different from the continuity equation examined so far.
This difference affects the density perturbations discussed in the next section.

However, we first consider the background evolution of the universe in the modified entropic-force model.
Combining Eq.\ (\ref{eq:FRW01(g)}) with Eq.\ (\ref{eq:FRW02(g)}), we obtain  
\begin{equation}
 \dot{H}    = -  C_{m} H^2    +   C_{4}  H_{0}^{2}    , 
\label{eq:dHC1C3hC4h(Cst)}
\end{equation}
where the dimensionless constants $C_{m}$ and $C_{4}$ are 
\begin{equation}
     C_{m} = 1.5   \quad   \textrm{and}  \quad  {C}_{4}  =  \frac{ 3 \alpha_{4} + 2 \gamma_{4}^{\textrm{irr}}  }{   2   }. 
\label{eq:Cm=1.5_C4}
\end{equation}
Here $C_{m} =1.5$ corresponds to a matter-dominated universe in the standard cosmology \cite{Weinberg1,Roy1}.
Solving Eq.\ (\ref{eq:dHC1C3hC4h(Cst)}), the evolution of the Hubble parameter \cite{Koma6} is given by 
\begin{equation}
 \left ( \frac{H} {H_{0}} \right )^{2}   
=  \left ( 1-  \frac{ C_{4} }{ C_{m} }  \right )   \left ( \frac{ a } {  a_{0} } \right )^{ -2 C_{m}}   +  \frac{ C_{4} }{ C_{m} }     , 
\label{eq:H/H0(C1C4)_00}
\end{equation}
where $a_{0}$ is the scale factor at the present time.
This solution is the same as in the standard $\Lambda$CDM model. 
That is, the constant term $C_{4}/C_{m}$ in Eq.\ (\ref{eq:H/H0(C1C4)_00}) behaves as if it were $\Omega_{\Lambda}$ in the standard $\Lambda$CDM model.
Similarly, $1- \frac{C_{4}}{C_{m}}$ behaves as if it were $\Omega_{m}$ in the standard $\Lambda$CDM model in a flat universe \cite{Koma6}. 
(Note that $\Omega_{m}$ and $\Omega_{\Lambda}$ represent the density parameter for matter and for $\Lambda$, respectively.
The density parameter for the radiation is neglected in order to focus attention on the late universe.)

We define the constant parameter 
\begin{equation}
 \tilde{\Omega}_{\Lambda} \equiv  \frac{ C_{4} }{ C_{m}  }  ,
\label{eq:tOmega_L}
\end{equation}
where $C_{4}$ and $C_{m}$ are given in Eq.\ (\ref{eq:Cm=1.5_C4}).
Using $ \tilde{\Omega}_{\Lambda} $ and $C_{m}=1.5$,  Eq.\ (\ref{eq:H/H0(C1C4)_00}) can be rearranged as 
\begin{equation}
 \left ( \frac{H} {H_{0}} \right )^{2}   =  \left ( 1-  \tilde{\Omega}_{\Lambda} \right )   \tilde{a}^{ - 3}   +  \tilde{\Omega}_{\Lambda}  \quad \textrm{where}  \quad   \tilde{a} = \frac{a} { a_{0}}  .
\label{eq:H/H0(C1C4)_02}
\end{equation}
Keep in mind that $ \tilde{\Omega}_{\Lambda} $ is \textit{not} the density parameter for $\Lambda$, but is instead a constant. 

In order to study a dissipative universe, we define a dissipation rate 
\begin{equation}
      \tilde{\mu}  \equiv  \frac{ \gamma_{4}^{\textrm{irr}} }{ C_{4}  }     =  \frac{ \tilde{\Omega}_{D} }{ \tilde{\Omega}_{\Lambda} }      ,
\label{eq:dissipation_0}
\end{equation}
where $\tilde{\Omega}_{D}$ is a constant parameter related to dissipative processes given by
\begin{equation}
      \tilde{\Omega}_{D}  \equiv  \frac{ \gamma_{4}^{\textrm{irr}} }{ C_{m} }      .
\label{eq:Omega_D_0}
\end{equation}
Note that $\gamma_{4}^{\textrm{irr}}$, which is given in Eq.\ (\ref{eq:beta4_GE_alpha4_delta}), is a dimensionless constant related to an irreversible entropy.
When $\gamma_{4}^{\textrm{irr}} = 0 $, we obtain $\tilde{\mu} =0$.
Accordingly, the present model is equivalent to the standard nondissipative $\Lambda$CDM model for which $\gamma_{4}^{\textrm{irr}} = 0 $ (so that $\tilde{\mu} =0$).
In contrast, when $\alpha_{4} =0$, one gets $\tilde{\mu} =1$ from Eq.\ (\ref{eq:dissipation_0}) because $C_{4} = \frac{ 3 \alpha_{4} + 2 \gamma_{4}^{\textrm{irr}}  }{2} = \gamma_{4}^{\textrm{irr}} $.
In this case, the present model is equivalent to the fully dissipative CCDM model proposed by Lima, Jesus, and Oliveira, abbreviated as the LJO model \cite{Lima2010,Lima2011}.
It is expected that investigating $0 \leq \tilde{\mu} \leq 1$ can bridge the gap between the $\Lambda$CDM and CCDM models.  
From Eq.\ (\ref{eq:dissipation_0}), $\tilde{\mu}$ is proportional to $\tilde{\Omega}_{D}$ when $ \tilde{\Omega}_{\Lambda} $ is fixed, as discussed in Sec.\ \ref{Results}.

In the present paper, the Hubble horizon is used as the preferred screen, because the apparent horizon coincides with the Hubble horizon in a spatially flat universe \cite{Easson1}.
If we instead consider a spatially nonflat universe, we would use the apparent horizon as the preferred screen \cite{Koma4,Koma5,Koma6}. 
The entropic-force model considered here differs from holographic dark energy models \cite{Pavon_2005}, even though a holographic principle \cite{Hooft-Bousso} is applied to both \cite{Koma6}.

\section{Density perturbations in the modified entropic-force model}
\label{Density perturbations}

This section analyzes the first-order density perturbations in the modified entropic-force model that includes constant entropic-force terms. 
This model is equivalent to the LJO model \cite{Lima2010,Lima2011} when $\alpha_{4} =0$.
Density perturbations in the LJO model have been examined using a neo-Newtonian approach \cite{Lima2011}. 
That approach was proposed by Lima \textit{et al.} \cite{Lima_Newtonian_1997}, following earlier ideas developed by McCrea \cite{McCrea_1951} and Harrison \cite{Harrison_1965} that attempted to describe a Newtonian universe having pressure \cite{Lima2011}. 
The present paper also uses a neo-Newtonian approach.

In previous work \cite{Koma6}, density perturbations were examined using that approach.
A perturbation analysis in cosmology generally requires a fully relativistic description \cite{Lima2011}; a nonrelativistic (Newtonian) approach only works when the scale of the perturbations is much less than the Hubble radius and the velocity of peculiar motions is small in comparison to the Hubble flow \cite{Lima2011}. 
However, such difficulties should be circumvented by the neo-Newtonian approximation \cite{Lima2011,Koma6}.

The modified Friedmann, acceleration, and continuity equations from Sec.\ \ref{Modified entropic-force model} are 
\begin{equation}
   H^{2}  = \frac{ 8\pi G }{ 3 } \rho    +  \alpha_{4}  H_{0}^{2}   , 
\label{eq:FRW01(g)_2}
\end{equation}
\begin{align} 
  \frac{ \ddot{a} }{ a }  &=  -  \frac{ 4\pi G }{ 3 } \rho    +   \alpha_{4}  H_{0}^{2}    +  \gamma_{4}^{\textrm{irr}} H_{0}^{2}                   \notag   \\
                                 &=  -  \frac{ 4\pi G }{ 3 } \left ( \rho  + \frac{3 p_{e} }{c^2}  \right ) + \alpha_{4}  H_{0}^{2}   , 
\label{eq:General_FRW02_g_2_alpha4_2}
\end{align}
and
\begin{equation}
       \dot{\rho} + 3  \frac{\dot{a}}{a}  \rho   =  \Gamma  \rho   , 
\label{eq:drho_General_01_alpha4_2}
\end{equation}
where the effective pressure $p_{e}$ from Eq.\ (\ref{eq:pe_Q_1_alpha4}) and $\Gamma$ from Eq.\ (\ref{eq:Gamma_0_alpha4}) are 
\begin{equation}
   p_{e} =  - \frac{  c^{2} H_{0}^{2}} {  4\pi G }   \gamma_{4}^{\textrm{irr}}  \quad \textrm{and}  \quad  \Gamma  = - 3 H \frac{ p_{e} }{\rho c^{2}}   , 
\label{eq:pe_Gamma_alpha4_2}
\end{equation}
assuming
\begin{equation}
   \alpha_{4}     \geq 0   \quad  \textrm{and}  \quad     \gamma_{4}^{\textrm{irr}}    \geq 0   .
\end{equation}
Using Eq.\ (\ref{eq:pe_Gamma_alpha4_2}), the effective equation of state parameter, $w_{e}$, becomes
\begin{equation}
  w_{e} = \frac{p_{e}}{\rho c^{2}}   =       - \frac{ H_{0}^{2} } {  4\pi G }\frac{ \gamma_{4}^{\textrm{irr}} }{\rho}   = - \frac{\Gamma}{3H}  .
\label{eq:we_p_alpha4}
\end{equation}
When $\alpha_{4} =0$, the above equations are identical to those in Refs.\ \cite{Lima_Newtonian_1997,Lima2011,Koma6}.

In general, $Q = - \frac{3}{8 \pi G}  \frac{ \dot{f}(t) }{  \rho  }$ from Eq.\ (\ref{eq:Q_0}) is nonzero.
Therefore, as shown in Eq. (\ref{eq:drho_General_01}), the right-hand side of the continuity equation includes not only $\Gamma \rho$ but also $Q \rho$.
However, $Q=0$ is obtained from $f(t) = \alpha_{4}$ in the present model.
That is, we can neglect the exchange of energy appearing in an energy-exchange cosmology \cite{C4}.
Consequently, the right-hand side of the continuity equation is $\Gamma \rho$, as shown by Eq. (\ref{eq:drho_General_01_alpha4_2}). 

In order to apply the neo-Newtonian approach, the basic hydrodynamical equations \cite{Lima_Newtonian_1997,Lima2011} for the modified entropic-force model are rewritten as 
\begin{equation}
 \left ( \frac{ \partial \mathbf{u} }{ \partial t } \right )_{r} + ( \mathbf{u} \cdot  \nabla_{r} ) \mathbf{u}   =  -  \nabla_{r} \Phi - \frac{ \nabla_{r} p_{e} } { \rho + \frac{ p_{e} }{ c^{2} } }   , 
\label{eq:Newtonian_1_alpha4}
\end{equation}
\begin{equation}
 \left ( \frac{ \partial \rho }{ \partial t } \right )_{r} +  \nabla_{r}  \cdot ( \rho \mathbf{u} ) + \frac{ p_{e} }{ c^{2} }  \nabla_{r} \cdot \mathbf{u}  =  0  , 
\label{eq:Newtonian_2_alpha4}
\end{equation}
and
\begin{equation}
\nabla_{r} ^{2} \Phi = 4 \pi G  \left ( \rho   +  l \right )   , 
\label{eq:Newtonian_3_alpha4}
\end{equation}
where $\mathbf{u}$ is the velocity of a volume fluid element and $\Phi$ is the gravitational potential.
In the present model,  $l$ is given by
\begin{equation} 
 l    =  \frac{ 3 p_{e} }{ c^{2} }  -  \frac{ 3 f(t) }{ 4 \pi G }  = \frac{ 3 p_{e} }{ c^{2} }  -  \frac{ 3 \alpha_{4}  H_{0}^{2}  }{ 4 \pi G }   , 
\label{eq:l_pe_alpha4}
\end{equation}
where the effective pressure is $p_{e}  = p + p_{c} = p_{c} $ in a matter-dominated universe (for which $p=0$).
Equations (\ref{eq:Newtonian_1_alpha4}) to (\ref{eq:Newtonian_3_alpha4}) are the Euler, continuity, and Poisson equations, respectively. 
The basic hydrodynamical equations are almost equivalent to those in Refs.\ \cite{Lima_Newtonian_1997,Lima2011,Koma6}.
However, in the present study, the Poisson equation is modified in order to take into account the $\alpha_{4} H_{0}^{2}$ terms corresponding to $\Lambda/3$.
For this purpose, the basic equations for the $\Lambda(t)$CDM models discussed in Ref. \cite{Waga1994} are adopted.
Consequently, Eq.\ (\ref{eq:l_pe_alpha4}) includes an $\alpha_{4} H_{0}^{2}$ term, slightly extending a previous formulation \cite{Koma6}.
As discussed in Ref.\ \cite{Sola_2009}, dark energy perturbations can be neglected in the $\Lambda(t)$CDM model.
This is justified in most cases \cite{Sola_2009,Sola_2007-2009}.
Similarly, it is assumed that perturbations in the $\alpha_{4} H_{0}^{2}$ terms are negligible in the present model.

Using the preceding equations, the time evolution equation for the matter density contrast, i.e., the perturbation growth factor $\delta \equiv \delta \rho_{m} /\rho_{m} $, can be calculated.
The derivation is essentially the same as that of Jesus \textit{et al.} \cite{Lima2011}. 
Setting $c=1$, using a linear approximation, and neglecting extra terms, we obtain the following time evolution equation for $\delta$,
\begin{align}
\ddot{\delta}  & + \left [ H (2  +  3 c_{\rm{eff}}^{2} - 3 w_{e} )   - \frac{ \dot{w_{e}} }{ 1+w_{e} }  \right ] \dot{\delta}        \notag \\
                     & +  \Bigg \{      3 (\dot{H} + 2 H^{2}) \left (   c_{\rm{eff}}^{2}  - w_{e} \right )     \notag \\
                     & + 3 H  \left [ \dot{c}_{\rm{eff}}^{2}   - ( 1 + c_{\rm{eff}}^{2} )  \frac{ \dot{w}_{e} }{ 1+w_{e}  } \right ] \notag \\     
                     &  - 4 \pi G \rho \left ( 1 + w_{e} \right ) ( 1 + 3 c_{\rm{eff}}^{2} )  + \frac{  k^{2} c_{\rm{eff}}^{2} }{ a^{2} }    \Bigg \}     \delta = 0              , 
\label{eq:delta-t_Unif_BV_0}
\end{align}
where the effective speed of sound is 
\begin{equation}
 c_{\rm{eff}}^{2}  \equiv   \frac{\delta p_{e} }{\delta \rho }   .
\label{eq:ceff_Unif_alpha4}
\end{equation}
Equation\ (\ref{eq:delta-t_Unif_BV_0}) is equivalent to one found in Ref.\ \cite{Koma6}. 
However, through $w_{e}$ and $\rho$, the above equation also implicitly includes $\alpha_{4} H_{0}^{2}$ terms.
In Eq.\ (\ref{eq:delta-t_Unif_BV_0}), $\rho_{m}$ is replaced by $\rho$ because only a single-fluid-dominated universe \cite{Koma6} is being considered.  
Also, $\rho$ in Eq.\ (\ref{eq:delta-t_Unif_BV_0}) represents the average value $\bar{\rho}$ corresponding to a homogenous and isotropic solution for the three unperturbed Friedmann, acceleration, and continuity equations.
For simplicity, set $c=1$ and replace $\bar{\rho}$ with $\rho$ when considering the time evolution equation for $\delta$ \cite{Koma6}.

As described in Ref.\ \cite{Lima2011}, assume that $c_{\rm{eff}}^{2} = c_{\rm{eff}}^{2}(t)$ and that the spatial dependence of $\delta$ is proportional to $e^{i \bf{k} \cdot \bf{x} }$, 
where the comoving coordinates $\bf{x}$ are related to the proper coordinates $\bf{r}$ by  $\bf{x} = \bf{r}$$/a$.
In addition, we assume $c_{\rm{eff}}^{2} = 0$ \cite{Koma6} because the neo-Newtonian equation is only equivalent to the general relativistic equation for a single-fluid-dominated universe when $c_{\rm{eff}}^{2} = 0$ \cite{Reis_2003}. 
That equivalence has been recently discussed in Ref.\ \cite{Ramos_2014}.
In the present model, the effective pressure ${p}_{e}$ is constant according to Eq.\ (\ref{eq:pe_Gamma_alpha4_2}).
Therefore, we find $c_{s}^{2} \equiv \dot{p}_{e}/\dot{\rho} = 0$ which
indicates adiabatic perturbations, $c_{\rm{eff}}^{2} = c_{s}^{2}$ \cite{Reis_2003,Ramos_2014}, since $c_{\rm{eff}}^{2} =0$.
The influence of $c_{\rm{eff}}^{2}$ has been examined in Ref.\ \cite{Lima2011}.
The case of $c_{\rm{eff}}^{2} \neq 0$ is not considered in the present study.

Substituting $ c_{\rm{eff}}^{2} =0$, $\dot{c}_{\rm{eff}}^{2} =0$, $w_{e}= - \frac{\Gamma}{3H}$,   
and $\frac{\dot{w}_{e} }{ 1+ w_{e} } =  \frac{ \Gamma \dot{H} - H \dot{\Gamma} }{  H  (3H - \Gamma) }$  into Eq.\ (\ref{eq:delta-t_Unif_BV_0}), we obtain
\begin{align}
\ddot{\delta}   + & \left [ 2 H + \Gamma  - \frac{ \Gamma \dot{H} - H \dot{\Gamma} }{ H (3H -\Gamma)}  \right ] \dot{\delta}        \notag \\
                      + & \Bigg \{        (\dot{H} + 2 H^{2})   \frac{ \Gamma }{ H }   - (3 H)  \frac{ \Gamma \dot{H} - H \dot{\Gamma} }{ H (3H -\Gamma)}  \notag \\
                         &  - 4 \pi G \rho \left ( 1 -  \frac{ \Gamma }{ 3H } \right )       \Bigg \}     \delta = 0              .
\label{eq:delta-t_CCDM_alpha4}
\end{align}
For numerical purposes, we define an independent variable \cite{Lima2011,Koma6} as
\begin{equation}
\eta \equiv \ln (\tilde{a}(t))   \quad \textrm{where} \quad  \tilde{a}(t) =  \frac{a(t)}{a_{0}}    .
\label{eta_def}
\end{equation}
Using Eq.\ (\ref{eta_def}), Eq.\ (\ref{eq:delta-t_CCDM_alpha4}) can be rearranged as 
\begin{equation}
\delta^{\prime \prime}  + F(\eta) \delta^{\prime}  +  G(\eta) \delta =0, 
\label{eq:delta-eta_c=0_CCDM_alpha4}
\end{equation}
where $F(\eta)$ and $G(\eta)$ are
\begin{equation}
F(\eta) =  2  +  \frac{ \Gamma + H^{\prime} }{ H }   -   \frac{  \Gamma H^{\prime} - H \Gamma^{\prime}  }{ H (3H -\Gamma) }   
\label{eq:F(eta)_0}
\end{equation}
and
\begin{align}
G(\eta)  =  &   \left (  \frac{ H^{\prime} }{ H }  + 2   \right )     \frac{ \Gamma }{ H }       -     \frac{ 3( \Gamma H^{\prime} - H \Gamma^{\prime} ) }{ H (3H -\Gamma) }   \notag \\
                &    - \frac{ 4 \pi G \rho }{ H^{2} } \left ( 1 -  \frac{ \Gamma }{ 3H } \right )    .
\label{eq:G(eta)_0}
\end{align}
A prime $^{\prime}$ represents a differential with respect to $\eta$, i.e., $d/d \eta$.
The mass density $\rho$ in a homogeneous, isotropic, and spatially flat universe is obtained from the modified Friedmann equation as
\begin{equation}
 \rho =    \frac{3}{8 \pi G}  (H^{2} - \alpha_{4}  H_{0}^{2} )             .        
\label{eq:rho_alpha4}
\end{equation}
The critical density $\rho_{c0}$ is
\begin{equation}
 \rho_{c0} =  \frac{3}{8 \pi G}  (H_{0}^{2} - \alpha_{4}  H_{0}^{2} )  =  \frac{3}{8 \pi G}  H_{0}^{2}  (1 - \alpha_{4} )             .        
\label{eq:rho_crirical}
\end{equation}

Next, from Eq.\ (\ref{eq:Gamma_0_alpha4}) or (\ref{eq:pe_Gamma_alpha4_2}), $\Gamma$ can be rewritten as
\begin{equation}
   \Gamma =   \frac{  3 H } {  4\pi G } \frac{ H_{0}^{2} \gamma_{4}^{\textrm{irr}} }{\rho}   .
\label{eq:Gamma_alpha4_3}
\end{equation}
Using the critical density $\rho_{c0}$ from Eq.\ (\ref{eq:rho_crirical}) and $C_{m}= 3/2$ from Eq.\ (\ref{eq:Cm=1.5_C4}), 
Eq.\ (\ref{eq:Gamma_alpha4_3}) can be rearranged as
\begin{align}
\Gamma & =  \frac{  3 H } {  4\pi G } \frac{ H_{0}^{2} \gamma_{4}^{\textrm{irr}} }{\rho}  \times \frac{ \rho_{c0} }{ \frac{3}{8 \pi G}  H_{0}^{2}  (1 - \alpha_{4} )   }                 \notag \\
             & =   \frac{ 3H \gamma_{4}^{\textrm{irr}} }{ \frac{3}{2} (1 - \alpha_{4}) }           \left (   \frac{ \rho_{c0}  }{ \rho }  \right )                                                         
                =   \frac{ 3   \gamma_{4}^{\textrm{irr}} }{ C_{m} (1 - \alpha_{4}) }            \left (   \frac{ \rho_{c0}  }{ \rho }  \right )            H                                              \notag \\
             & =   \frac{ 3 \tilde{\Omega}_{D} }{ 1 - \alpha_{4}} \left (   \frac{ \rho_{c0}  }{ \rho }  \right ) H             ,
\label{eq:gamma_Omega_alpha4_0}
\end{align}
where $\tilde{\Omega}_{D}$ from Eq.\ (\ref{eq:Omega_D_0}) is 
\begin{equation}
      \tilde{\Omega}_{D}  \equiv  \frac{ \gamma_{4}^{\textrm{irr}} }{ C_{m} }      .
\label{eq:Omega_D}
\end{equation}
From Eqs.\ (\ref{eq:Cm=1.5_C4}),  (\ref{eq:tOmega_L}), and (\ref{eq:Omega_D}), $\tilde{\Omega}_{\Lambda} -\tilde{\Omega}_{D}$ becomes 
\begin{align}
    \tilde{\Omega}_{\Lambda} -\tilde{\Omega}_{D}  &=   \frac{C_{4}}{C_{m}}   -  \frac{ \gamma_{4}^{\textrm{irr}}  }{C_{m}}   
                                                                         =   \frac{ \frac{ 3 \alpha_{4} + 2 \gamma_{4}^{\textrm{irr}} }{2} - \gamma_{4}^{\textrm{irr}}  }{ 3/2 }    \notag \\
                                                                      &=     \alpha_{4}  .
\label{eq:Omega_L-Omega_D}
\end{align}
Substituting Eqs.\ (\ref{eq:rho_alpha4}) and (\ref{eq:rho_crirical}) into Eq.\ (\ref{eq:gamma_Omega_alpha4_0}), replacing $(H/H_{0})^{2}$ by Eq.\ (\ref{eq:H/H0(C1C4)_02}), and using Eq.\ (\ref{eq:Omega_L-Omega_D}) and $\tilde{a} = e^{\eta}$, we find
\begin{align}
\frac{\Gamma}{H} &=   \frac{ 3 \tilde{\Omega}_{D} }{ 1 - \alpha_{4}} \left (   \frac{ \rho_{c0}  }{ \rho }  \right )   
                             =   \frac{ 3 \tilde{\Omega}_{D} }{ 1 - \alpha_{4}} \left (  \frac{   H_{0}^{2} - \alpha_{4}  H_{0}^{2} }{  H^{2} - \alpha_{4}  H_{0}^{2}}     \right )                             \notag \\
                           &=   \frac{ 3 \tilde{\Omega}_{D} }{   (H/H_{0})^{2} - \alpha_{4}  }                                              
                             = \frac{ 3 \tilde{\Omega}_{D} }{  ( 1-  \tilde{\Omega}_{\Lambda}  )   \tilde{a}^{ - 3}   +  \tilde{\Omega}_{\Lambda}        - \alpha_{4}  }                                    \notag \\
                           &= \frac{ 3 \tilde{\Omega}_{D} \tilde{a}^{3}  }{  ( 1-  \tilde{\Omega}_{\Lambda}  )  +  (\tilde{\Omega}_{\Lambda}        - \alpha_{4})   \tilde{a}^{3}            }         \notag \\                
                           &= \frac{ 3 \tilde{\Omega}_{D}  e^{3 \eta}    }{    1-  \tilde{\Omega}_{\Lambda}     +  \tilde{\Omega}_{D}  e^{ 3 \eta}                                                     }    .
\label{eq:Gamma-H_alpha4_1}
\end{align}
Equation\ (\ref{eq:Gamma-H_alpha4_1}) includes not only $\tilde{\Omega}_{\Lambda}$ but also $\tilde{\Omega}_{D}$.
Similarly, we obtain
\begin{equation}
      \frac{\Gamma + H^{\prime} }{H} = \frac{   3 \tilde{\Omega}_{D}  e^{3 \eta}    }{    1-  \tilde{\Omega}_{\Lambda}     +  \tilde{\Omega}_{D}  e^{ 3 \eta}   }   +  \frac{ -\frac{3}{2} ( 1- \tilde{\Omega}_{\Lambda} )   }{   1-  \tilde{\Omega}_{\Lambda}      +  \tilde{\Omega}_{\Lambda}   e^{3 \eta}              }    ,
\label{eq:Gamma-H_alpha4_2}
\end{equation}
%
\begin{equation}
 \frac{  \Gamma H^{\prime} - H \Gamma^{\prime}  }{ H (3H -\Gamma) }  
      =    \frac{ - 3  \tilde{\Omega}_{D}  e^{ 3 \eta}  }{   1-  \tilde{\Omega}_{\Lambda}  +  \tilde{\Omega}_{D}   e^{ 3 \eta}              }    , 
\label{eq:Gamma-H_alpha4_3}
\end{equation}
\begin{equation}
    \frac{ 4 \pi G \rho }{ H^{2} } 
       =   \frac{3}{2}   \left (  1    - \frac{  (\tilde{\Omega}_{\Lambda}  - \tilde{\Omega}_{D})  e^{ 3 \eta}  }{   1-  \tilde{\Omega}_{\Lambda}  +  \tilde{\Omega}_{\Lambda}   e^{ 3 \eta}              }   \right )     .
\label{eq:Gamma-H_alpha4_4}
\end{equation}
Substitution of Eqs.\ (\ref{eq:Gamma-H_alpha4_2}) and (\ref{eq:Gamma-H_alpha4_3}) into Eq.\ (\ref{eq:F(eta)_0}) results in
%
\begin{equation}
    F(\eta)=    2     +   \frac{   6 \tilde{\Omega}_{D}  e^{3 \eta}    }{    1-  \tilde{\Omega}_{\Lambda}     +  \tilde{\Omega}_{D}  e^{ 3 \eta}   }   -  \frac{ 3 ( 1- \tilde{\Omega}_{\Lambda} )  }{ 2 (  1-  \tilde{\Omega}_{\Lambda}      +  \tilde{\Omega}_{\Lambda}   e^{3 \eta}    )         }    .
\label{eq:F_alpha4_OmegaL-Omegad_01}
\end{equation}
Likewise, substituting Eqs.\ (\ref{eq:Gamma-H_alpha4_1}) to (\ref{eq:Gamma-H_alpha4_4}) into Eq.\ (\ref{eq:G(eta)_0}) leads to
%
%
\begin{align}
 G(\eta) &=      \frac{   3   }{  2 ( 1-  \tilde{\Omega}_{\Lambda}     +  \tilde{\Omega}_{D}  e^{ 3 \eta}  ) (  1-  \tilde{\Omega}_{\Lambda}      +  \tilde{\Omega}_{\Lambda}   e^{3 \eta}   ) }                                                   \notag \\
             &\times    \left [  10 \tilde{\Omega}_{D}  \tilde{\Omega}_{\Lambda}  e^{6 \eta}     + 6 \tilde{\Omega}_{D} ( 1-   \tilde{\Omega}_{\Lambda}  ) e^{3 \eta}    -   \left ( 1 - \tilde{\Omega}_{\Lambda} \right )^{2}    \right ]    .
\label{eq:G_alpha4_OmegaL-Omegad_01}
\end{align}
As Eqs.\ (\ref{eq:F_alpha4_OmegaL-Omegad_01}) and (\ref{eq:G_alpha4_OmegaL-Omegad_01}) indicate, $F(\eta)$ and $G(\eta)$ include both $\tilde{\Omega}_{\Lambda}$ and $\tilde{\Omega}_{D}$.
From Eqs.\ (\ref{eq:tOmega_L}) and (\ref{eq:Omega_D}), they can be written as 
\begin{equation}
    \tilde{\Omega}_{\Lambda} =   \frac{C_{4}}{C_{m}}      \quad   \textrm{and}   \quad  \tilde{\Omega}_{D}  =   \frac{ \gamma_{4}^{\textrm{irr}} }{C_{m}}   ,
\label{eq:Omega_L_and_Omega_D}
\end{equation}
where $C_{m} = 1.5$ and $C_{4} = \frac{ 3 \alpha_{4} + 2 \gamma_{4}^{\textrm{irr}} }{2}$ according to Eq.\ (\ref{eq:Cm=1.5_C4}).
Here, $ \tilde{\Omega}_{\Lambda} $ is \textit{not} the density parameter for $\Lambda$, but is instead a constant, even though $ \tilde{\Omega}_{\Lambda} $ behaves as if it were $\Omega_{\Lambda}$ in the standard $\Lambda$CDM model.
In contrast, $\tilde{\Omega}_{D}$ is a constant parameter related to a dissipative process.
For confirmation, we consider two typical cases.
When $\gamma_{4}^{\textrm{irr}} =0$, one has $\tilde{\Omega}_{D}=0$. 
Substituting $\tilde{\Omega}_{D}=0$ into Eqs.\ (\ref{eq:F_alpha4_OmegaL-Omegad_01}) and (\ref{eq:G_alpha4_OmegaL-Omegad_01}), $F(\eta)$ and $G(\eta)$ recover the values of the $\Lambda (t)$-$C_\textrm{cst}$ model \cite{Koma6} corresponding to the standard $\Lambda$CDM model.
In contrast, when $\alpha_{4} =0$, we find that $\tilde{\Omega}_{D}= \tilde{\Omega}_{\Lambda} = C_{4}/C_{m}$ because $C_{4} = \gamma_{4}^{\textrm{irr}} $ results from Eq.\ (\ref{eq:Cm=1.5_C4}).
Therefore, when $\tilde{\Omega}_{D}= \tilde{\Omega}_{\Lambda}$,  $F(\eta)$ and $G(\eta)$ reduce to the expressions in the BV-$C_\textrm{cst}$ model \cite{Koma6} corresponding to the LJO model \cite{Lima2011}.

In the present paper, the differential equation is numerically solved for the matter density contrast $\delta$ in Eq.\ (\ref{eq:delta-eta_c=0_CCDM_alpha4}).
For this purpose, the initial conditions of the Einstein--de Sitter growing model \cite{Lima2011} are used. 
The initial conditions are taken to be $\delta (\tilde{a}_{i}) = \tilde{a}_{i}$ and $\delta^{\prime}  (\tilde{a}_{i}) = \tilde{a}_{i}$, where $\tilde{a}_{i} =a_{i}/a_{0} = 10^{-3}$ \cite{Koma6}.

\section{Influence of the dissipation rate}
\label{Results}

In this section, the influence of a dissipation rate in the modified entropic-force model is analyzed. 
We first consider the following related parameters.
According to Eq.\ (\ref{eq:dissipation_0}), the dissipation rate $\tilde{\mu}$ is
\begin{equation}
      \tilde{\mu}  \equiv  \frac{ \gamma_{4}^{\textrm{irr}} }{ C_{4}  }     =  \frac{ \tilde{\Omega}_{D} }{ \tilde{\Omega}_{\Lambda} }      ,
\label{eq:dissipation_1}
\end{equation}
where $\tilde{\Omega}_{\Lambda}$ and $\tilde{\Omega}_{D}$ from Eq.\ (\ref{eq:Omega_L_and_Omega_D}) are 
\begin{equation}
    \tilde{\Omega}_{\Lambda} =   \frac{C_{4}}{C_{m}}   \quad   \textrm{and}   \quad  \tilde{\Omega}_{D}  =   \frac{ \gamma_{4}^{\textrm{irr}} }{C_{m}}   ,  
\label{eq:Omega_L_and_Omega_D_2}
\end{equation}
and $C_{m}$ and $C_{4}$ from Eq.\ (\ref{eq:Cm=1.5_C4}) are 
\begin{equation}
     C_{m} = 1.5      \quad   \textrm{and}   \quad   C_{4} = \frac{ 3 \alpha_{4} + 2 \gamma_{4}^{\textrm{irr}}  }{2}     .
\label{eq:Cm_C4_02}
\end{equation}
When $\alpha_{4} =0$ (as in the CCDM models), we obtain $\tilde{\mu} =1$, whereas $\tilde{\mu} =0$ when $ \gamma_{4}^{\textrm{irr}}=0$ (corresponding to $\Lambda$CDM models).
To examine the influence of the dissipation rate, $\tilde{\mu}$ can be varied between $0$ and $1$.

Here, $\tilde{\Omega}_{\Lambda}$ is determined from the background evolution of the universe.
To this end, we identify $\tilde{\Omega}_{\Lambda}$ with $\Omega_{\Lambda}$ from a fine-tuned standard $\Lambda$CDM model \cite{Koma5,Koma6}.
In the standard $\Lambda$CDM model, we consider a spatially flat universe in which $(\Omega_{m}, \Omega_{\Lambda}) = (0.315, 0.685)$ based on the Planck 2013 results \cite{Planck2013}.
That is, set $\tilde{\Omega}_{\Lambda} =\Omega_{\Lambda}  = 0.685$.
To determine the influence of the dissipation rate, $\tilde{\mu}$ is set to several typical values, $0$, $0.05$, $0.1$, $0.2$, $0.4$, $0.6$, $0.8$, and $1.0$ in turn.
The background evolution of the universe for each case is equivalent to that in the standard $\Lambda$CDM model because $\tilde{\Omega}_{\Lambda} =\Omega_{\Lambda}  = 0.685$.
Accordingly, every case agrees with the observed supernova data. 
We calculate $\alpha_4$ and $\gamma_{4}^{\textrm{irr}}$ from $\tilde{\Omega}_{\Lambda}  = 0.685$ and the preceding value of $\tilde{\mu}$, using $\alpha_{4} = (1 - \tilde{\mu}) \tilde{\Omega}_{\Lambda}$ and $\gamma_{4}^{\textrm{irr}}  = 3 \tilde{\mu} \tilde{\Omega}_{\Lambda} /2$.
Substituting $\tilde{\Omega}_{\Lambda}  = 0.685$ into Eq.\ (\ref{eq:dissipation_1}), we find $\tilde{\Omega}_{D} = 0.685 \tilde{\mu}$.

\begin{figure} [t] 
\begin{minipage}{0.495\textwidth}
\begin{center}
\scalebox{0.3}{\includegraphics{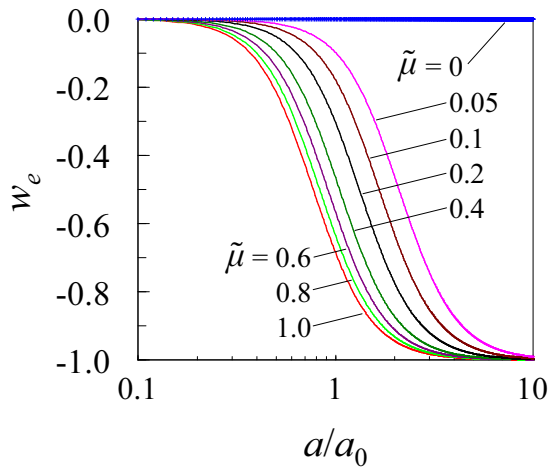}}
\end{center}
\end{minipage}
\caption{ (Color online). Evolution of the effective equation of state parameter $w_{e}$ for the indicated dissipation rates $\tilde{\mu}$. 
Note that $\tilde{\mu} =0$ corresponds to $\Lambda$CDM models, whereas $\tilde{\mu} =1$ corresponds to CCDM models.
The background evolution of the universe in each case is equivalent to that in the standard $\Lambda$CDM model because $\tilde{\Omega}_{\Lambda}  = \Omega_{\Lambda} = 0.685$.      }
\label{Fig-we-a}
\end{figure}

We consider the effective equation of state parameter $w_{e}$ in Eq.\ (\ref{eq:we_p_alpha4}). 
Substituting Eq.\ (\ref{eq:Gamma-H_alpha4_1}) into that equation, and replacing $e^{3 \eta}$ by $\tilde{a}^{3}$, we obtain
\begin{equation}
w_{e} = - \frac{  \tilde{\Omega}_{D}  \tilde{a}^{3}    }{    1-  \tilde{\Omega}_{\Lambda}     +  \tilde{\Omega}_{D}  \tilde{a}^{3}                            }    ,
\label{eq:we-a_alpha4}
\end{equation}
where $\tilde{a}$ is the normalized scale factor $a/a_{0}$.
Using this result, the evolution of $w_{e}$ can be determined as a function of the dissipation rate. 
As shown in Fig.\ \ref{Fig-we-a}, $w_{e}$ for $\tilde{\mu} =0$ is equal to $0$, because the effective pressure $p_{e} = 0$.
However,  $w_{e}$ for $\tilde{\mu} >0$ gradually decreases with increasing $a/a_{0}$ and finally approaches $-1$.
In addition, $w_{e}$ decreases with increasing $\tilde{\mu}$.
The dissipation rate $\tilde{\mu}$ thereby affects $w_{e}$ even if the background evolution of the universe is equivalent to that in the standard $\Lambda$CDM model.
The equation of state parameter for a generic component of matter, i.e., $w = p/(\rho c^2)$, is always zero in a matter-dominated universe (for which $p=0$).
That is, $w$ is not equal to $w_{e}$.
(Note that a generalized inhomogeneous equation of state has been discussed in Ref.\ \cite{Odintsov_2006_b}.)

\begin{figure} [t] 
\begin{minipage}{0.495\textwidth}
\begin{center}
 \scalebox{0.3}{\includegraphics{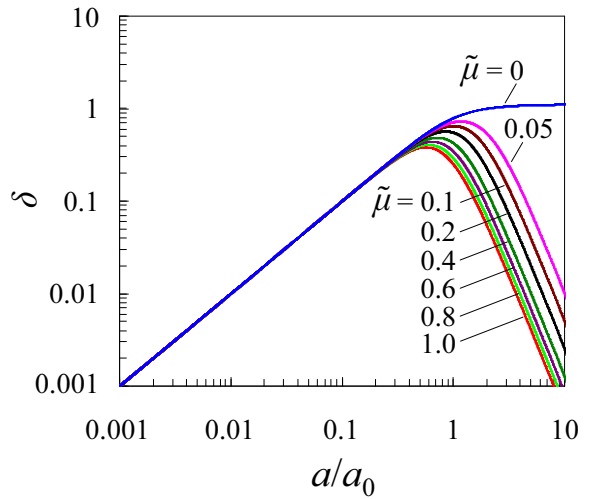}}
\end{center}
\end{minipage}
\caption{ (Color online). Evolution of the density perturbation growth factor $\delta$ for various dissipation rates $\tilde{\mu}$. 
The initial conditions are $\delta (\tilde{a}_{i}) = \tilde{a}_{i}$ and $\delta^{\prime}  (\tilde{a}_{i}) = \tilde{a}_{i}$ where $\tilde{a}_{i} =a_{i}/a_{0} = 10^{-3}$.  
The background evolution of the universe is the same in every case because $\tilde{\Omega}_{\Lambda} = 0.685$.      }
\label{Fig-delta-a}
\end{figure}

We next consider the first-order density perturbations in the model for various values of $\tilde{\mu}$.
It is helpful to calculate the evolution of the perturbation growth factor $\delta$.
As shown in Fig.\ \ref{Fig-delta-a}, $\delta$ increases with $a/a_{0}$  when $a/a_{0} \lessapprox 0.1$.
In contrast, when $a/a_{0} \gtrapprox 0.1$, $\delta$ increases less and eventually turns around and decreases, except for the case of $\tilde{\mu} =0$ (corresponding to the standard $\Lambda$CDM model).
Consequently, with increasing $\tilde{\mu}$, the perturbation growth factor deviates from $\delta$ for $\tilde{\mu} =0$. 
In that way, the dissipation rate $\tilde{\mu}$ affects the density perturbations.
However, $\delta$ for $\tilde{\mu}=0.05$ and $0.1$ is not much different from $\delta$ for $\tilde{\mu}=0$ at the present time when $a/a_{0} \approx 1$.
To study this effect more closely, we consider the growth rate for clustering \cite{Peebles_1993}.
That growth rate has been previously examined in the $\Lambda(t)$CDM and CCDM models \ \cite{Sola_2009,Lima2011}.

\begin{figure} [t] 
\begin{minipage}{0.495\textwidth}
\begin{center}
\scalebox{0.32}{\includegraphics{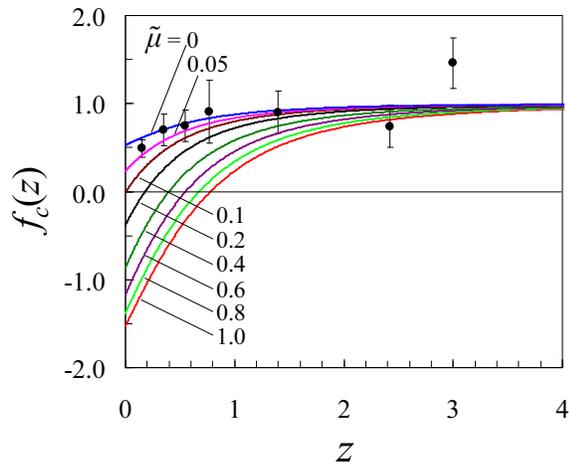}}
\end{center}
\end{minipage}
\caption{ (Color online). Evolution of the growth rate $f_{c}(z)$ for clustering for various dissipation rates $\tilde{\mu}$. 
The closed circles with error bars are the observed data points summarized in Ref.\ \cite{Lima2011}. 
The original data are from Refs.\ \cite{Colless2001,Guzzo2008,Tegmark2006,Ross2007,Angela2007,Viel2004,McDonald2005}. 
The background evolution of the universe for each value of $\tilde{\mu}$ is the same because $\tilde{\Omega}_{\Lambda}  = 0.685$.     }
\label{Fig-f(z)-z}
\end{figure}

The growth rate $f_{c}(z)$ for clustering is
\begin{equation}
 f_{c}(z) = \frac{d \ln \delta }{  d \ln a } = - (1 + z ) \frac{d \ln \delta }{  dz }     , 
\label{eq:f(z)}
\end{equation}
where  the redshift $z$ is 
\begin{equation}
   z \equiv \frac{ a_0 }{ a } -1   . 
\end{equation}
Keep in mind that $f_{c}(z)$ is not the extra driving term $f(t)$ shown in Eq.\ (\ref{eq:General_FRW01_f_0}).
The evolution of the growth rate is plotted in Fig.\ \ref{Fig-f(z)-z}.
The observed data points are taken from the summary in Ref.\ \cite{Lima2011}.
As shown in Fig.\ \ref{Fig-f(z)-z}, for large redshifts ($z \gtrapprox 2$), the calculated value of $f_{c}(z)$ is positive and consistent with the observations. 
However, for low redshifts ($z \lessapprox 1$), the theoretical expression for $f_{c}(z)$ deviates from the observed data points. 
This deviation occurs because, as shown in Fig.\ \ref{Fig-delta-a}, the calculated value of $\delta$ decays at high $a/a_{0}$ (corresponding to low $z$). 
In particular, $f_{c}(z)$ for $\tilde{\mu} = 1$, which corresponds to the CCDM models, markedly deviates from the observed points at low $z$, consistent with the results in previous works \cite{Lima2011,Koma6,Ramos_2014}.
However, $f_{c}(z)$ for $\tilde{\mu} = 1$ agrees with observations if $c_{\rm{eff}}^{2}$ is equal to $-1$ which makes sense only if $c_{\rm{eff}}^{2}$ is a free parameter, as in Ref.\ \cite{Lima2011}.
Further, the calculated results are expected to agree with the observed data for \textit{clustered matter} \cite{Ramos_2014}.

\begin{figure} [t] 
\begin{minipage}{0.495\textwidth}
\begin{center}
 \scalebox{0.31}{\includegraphics{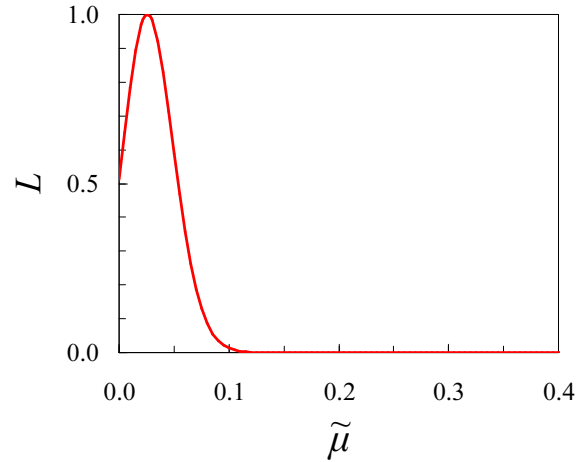}}
\end{center}
\end{minipage}
\caption{ (Color online). The normalized likelihood $L$ as a function of the dissipation rate $\tilde{\mu}$. A pure $\Lambda$CDM model corresponds to $\tilde{\mu} =0$. The maximum value of $L$ (corresponding to minimum $\chi^{2}$) is obtained for $\tilde{\mu} = 0.026$, upon sampling $\tilde{\mu} \in [0.020, 0.030]$ in steps of $0.001$, with $\tilde{\Omega}_{\Lambda} = 0.685$.    }
\label{Fig-L-mu}
\end{figure}

As graphed in Fig.\ \ref{Fig-f(z)-z},  $f_{c}(z)$ agrees well with the observed data points not only for $\tilde{\mu} = 0$ but also for $\tilde{\mu} = 0.05$ and $0.1$.
This agreement implies that low dissipation rates describe structure formation. 
To confirm this conclusion, a likelihood analysis is performed.
Here, the dissipation rate $\tilde{\mu}$ is a free parameter because $\tilde{\Omega}_{\Lambda} = 0.685$.
Accordingly, the chi-squared function used in Ref.\ \cite{Lima2011} can be rewritten as 
\begin{equation}
\chi^{2} (\tilde{\mu}) = \sum\limits_{i=1}^{7} { \left[ \frac{   f_{c}^{\textrm{obs}} (z_{i})  - f_{c}^{\textrm{cal}} (z_{i}, \tilde{\mu} )   }{ \sigma_{i} }   \right]^{2}   }   ,
\label{eq:chi_mu}
\end{equation}
where $f_{c}^{\textrm{obs}} (z)$ and $f_{c}^{\textrm{cal}} (z, \tilde{\mu} )$ are the observed and calculated growth rates, respectively, and $\sigma$ is the uncertainty in the observed growth rate.
The seven observed data points (numbered $i=1$ to $7$) shown in Fig.\ \ref{Fig-f(z)-z} are taken from the summary in Ref.\ \cite{Lima2011}. 
For the likelihood analysis, $\tilde{\mu}$ is sampled in the range $[0, 0.4]$ in steps of $0.005$.
Negative dissipation rates have not been considered.

A likelihood function $L$ \cite{Lima2010} is calculated as 
\begin{equation}
 L \propto \exp ({- \chi^{2}/ 2}) . 
\label{eq:L-chi}
\end{equation}
This equation indicates that high $L$ corresponds to low $\chi^{2}$ and vice versa.
For simplicity, the likelihood function $L$ is normalized below.
Figure\ \ref{Fig-L-mu} plots the normalized likelihood function $L$ for increasing $\tilde{\mu}$. 
It can be seen that $L$ is large for low dissipation rates, $\tilde{\mu} \lessapprox 0.1$.
Such a low-dissipation model agrees well with observation. 
This result suggests a weakly dissipative universe.
As illustrated in Fig.\ \ref{Fig-f(z)-z}, a low dissipation rate ($0 < \tilde{\mu} \lessapprox 0.1$) predicts a smaller growth rate than that in the standard pure $\Lambda$CDM model (for which $\tilde{\mu} = 0$).
Hopefully, future more detailed observations will be able to distinguish a low-dissipation model from a pure $\Lambda$CDM one.

\begin{figure} [t] 
\begin{minipage}{0.495\textwidth}
\begin{center}
 \scalebox{0.3}{\includegraphics{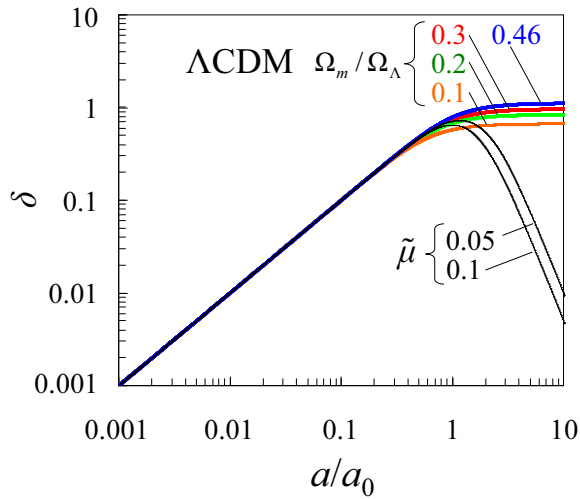}}
\end{center}
\end{minipage}
\caption{ (Color online). Evolution of the density perturbation growth factor $\delta$ for various values of $\Omega_{m} / \Omega_{\Lambda} $ in a pure $\Lambda$CDM model. 
Bold lines (in color) represent the pure $\Lambda$CDM model (for which $\tilde{\mu} = 0$), whereas (black) thin lines represent a low-dissipation model, i.e., $\tilde{\mu} = 0.05$ and $0.1$ with $\tilde{\Omega}_{\Lambda} = 0.685$.
In the pure $\Lambda$CDM model, $\Omega_{m} / \Omega_{\Lambda}  = 0.1$, $0.2$, $0.3$, and $0.46$ are approximately equivalent to $\Omega_{\Lambda} = 0.909$, $0.833$, $0.769$, and $0.685$, respectively. 
$\Omega_{m} / \Omega_{\Lambda} = 0.46$ corresponds to a fine-tuned standard $\Lambda$CDM model, which is equivalent to $\tilde{\mu} = 0$ shown in Fig.\ \ref{Fig-delta-a}.  
The background evolution of the universe for each value of $\Omega_{m} / \Omega_{\Lambda}$ is different because $\Omega_{\Lambda}$ is different. 
In contrast, the background evolutions of the universe for $\Omega_{m} / \Omega_{\Lambda} = 0.46$, $\tilde{\mu} = 0.05$, and $\tilde{\mu} = 0.1$ are the same because  $\Omega_{\Lambda} = \tilde{\Omega}_{\Lambda} = 0.685$. }
\label{Fig-delta-a_Lambda}
\end{figure}

\begin{figure} [t] 
\begin{minipage}{0.495\textwidth}
\begin{center}
\scalebox{0.32}{\includegraphics{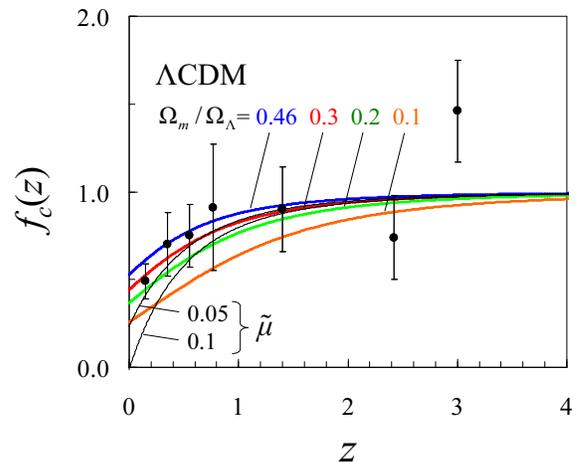}}
\end{center}
\end{minipage}
\caption{ (Color online). Evolution of the growth rate $f_{c}(z)$ for clustering for various values of $\Omega_{m} / \Omega_{\Lambda} $ in a pure $\Lambda$CDM model. 
The closed circles with error bars are the observed data points summarized in Ref.\ \cite{Lima2011}.  
$\Omega_{m} / \Omega_{\Lambda} = 0.46$ (i.e., $\Omega_{\Lambda} =  0.685$) corresponds to a fine-tuned standard $\Lambda$CDM model, which is equivalent to $\tilde{\mu} = 0$ shown in Fig.\ \ref{Fig-f(z)-z}. 
The background evolution of the universe for each value of $\Omega_{m} / \Omega_{\Lambda}$ is different.
See the caption of Fig.\ \ref{Fig-delta-a_Lambda}.   }
\label{Fig-f(z)-z_Lambda}
\end{figure}

\begin{figure} [t] 
\begin{minipage}{0.495\textwidth}
\begin{center}
\scalebox{0.33}{\includegraphics{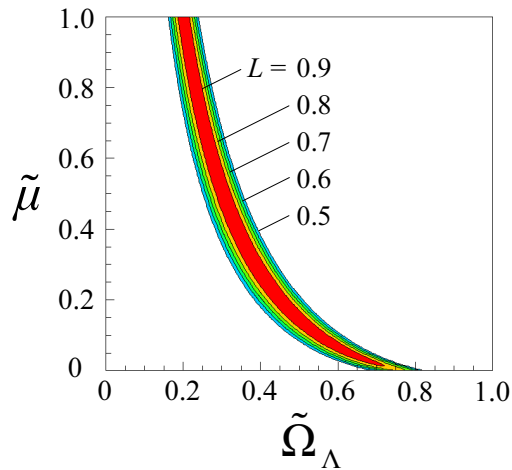}}
\end{center}
\end{minipage}
\caption{ (Color online). The contours of the normalized likelihood $L$ in the $(\tilde{\Omega}_{\Lambda}, \tilde{\mu})$ plane. 
To calculate a likelihood function in the $(\tilde{\Omega}_{\Lambda}, \tilde{\mu})$ plane,  Eqs.\ (\ref{eq:chi_mu}) and (\ref{eq:L-chi}) are used.
For this purpose, $\chi^{2} (\tilde{\mu})$ and $f_{c}^{\textrm{cal}} (z_{i}, \tilde{\mu} )$ in Eq.\ (\ref{eq:chi_mu}) are replaced by $\chi^{2} (\tilde{\Omega}_{\Lambda}, \tilde{\mu})$ and $f_{c}^{\textrm{cal}} (z_{i}, \tilde{\Omega}_{\Lambda}, \tilde{\mu} )$, respectively. 
For the likelihood analysis, $\tilde{\Omega}_{\Lambda}$ and $\tilde{\mu}$ are sampled in the range $[0, 1]$ in steps of $0.005$.
The likelihood function is normalized, using the maximum value which is obtained for $(\tilde{\Omega}_{\Lambda}, \tilde{\mu}) = (0.250, 0.710)$.
The contours of $L=0.9$, $0.8$, $0.7$, $0.6$, and $0.5$ are plotted.  }
\label{Fig--L_mu_Lambda}
\end{figure}

The background evolution of the universe considered so far is the same in every case because $\tilde{\Omega}_{\Lambda} = \Omega_{\Lambda} = 0.685$.
Consequently, a low-dissipation model ($0 < \tilde{\mu} \lessapprox 0.1$) is found to be better than the standard pure $\Lambda$CDM model (for which $\tilde{\mu} = 0$).
However, a pure $\Lambda$CDM model for $\Omega_{\Lambda} > 0.685$ may be equivalent to the low-dissipation model for $\tilde{\Omega}_{\Lambda} = 0.685$.
Accordingly, we examine the influence of $\Omega_{\Lambda}$ in the pure $\Lambda$CDM model.
To this end, we consider a parameter $\Omega_{m} / \Omega_{\Lambda}$, where $\Omega_{m}$ is given by $\Omega_{m} = 1 - \Omega_{\Lambda}$ in a spatially flat universe.
For the pure $\Lambda$CDM model, $\Omega_{m} / \Omega_{\Lambda}$ is set several typical values, $0.1$, $0.2$, $0.3$, and $0.46$ in turn, which are approximately equivalent to $\Omega_{\Lambda} = 0.909$, $0.833$, $0.769$, and $0.685$, respectively.
Thus, the background evolution of the universe for each value of $\Omega_{m} / \Omega_{\Lambda}$ is different because $\Omega_{\Lambda}$ is different. 
Here, we calculate $\Omega_{\Lambda}$ from the preceding value of $\Omega_{m} / \Omega_{\Lambda}$, using $ \Omega_{\Lambda} = 1 / (1+ \Omega_{m} / \Omega_{\Lambda} )$ in a spatially flat universe.

We now observe the evolutions of the density perturbation growth factor $\delta$ and the growth rate $f_{c}(z)$, for various values of $\Omega_{m} / \Omega_{\Lambda} $ in a pure $\Lambda$CDM model.
In Figs.\ \ref{Fig-delta-a_Lambda} and \ref{Fig-f(z)-z_Lambda},  (color) bold lines represent the pure $\Lambda$CDM model  (for which $\tilde{\mu} = 0$), whereas (black) thin lines represent a low-dissipation model for $\tilde{\Omega}_{\Lambda} = 0.685$.
For the low-dissipation model, the results of $\tilde{\mu} = 0.05$ and $0.1$ shown in Figs.\ \ref{Fig-delta-a} and \ref{Fig-f(z)-z} are replotted in Figs.\ \ref{Fig-delta-a_Lambda} and \ref{Fig-f(z)-z_Lambda}.
Also, $\Omega_{m} / \Omega_{\Lambda} = 0.46$ corresponds to a fine-tuned standard $\Lambda$CDM model, which is equivalent to $\tilde{\mu} = 0$ shown in Figs.\ \ref{Fig-delta-a} and \ref{Fig-f(z)-z}.
As graphed in Fig.\ \ref{Fig-delta-a_Lambda}, $\delta$ for each $\Omega_{m} / \Omega_{\Lambda} $ in the pure $\Lambda$CDM model (for which $\tilde{\mu} =0$) increases with $a/a_{0}$ when $a/a_{0} \lessapprox 0.1$.
Thereafter, each curve for the pure $\Lambda$CDM model gradually tends to a gentle incline.
Consequently, when $a/a_{0} \gg 1$, $\delta$ for the $\Lambda$CDM model is larger than $\delta$ for $\tilde{\mu} =0.05$ and $0.1$ in the low-dissipation model.
However, $\delta$ for the $\Lambda$CDM model is not much different from $\delta$ for the low-dissipation model when $a/a_{0} \lessapprox 1$.
To examine this effect more closely, we observe the growth rate $f_{c}(z)$ for clustering.
As shown in Fig.\ \ref{Fig-f(z)-z_Lambda}, $f_{c}(z)$ for $\Omega_{m} / \Omega_{\Lambda} = 0.3$ and $0.2$ in the pure $\Lambda$CDM model is likely consistent with $f_{c}(z)$ for $\tilde{\mu} =0.05$ and $0.1$ in the low-dissipation model, respectively.
That is, when $z \geq 0$ (i.e., $a/a_{0} \leq 1$), density perturbations in a pure $\Lambda$CDM model for $0.7 \lessapprox \Omega_{\Lambda} \lessapprox 0.8$ are similar to those in a low-dissipation model for $\tilde{\Omega}_{\Lambda} = 0.685$. 
(It should be noted that $\Omega_{m} / \Omega_{\Lambda} = 0.2$, $0.3$, and $0.46$ correspond to $\Omega_{\Lambda} = 0.833$, $0.769$, and $0.685$, respectively.) 
To confirm this conclusion, we observe the contours of the normalized likelihood $L$ in the $(\tilde{\Omega}_{\Lambda}, \tilde{\mu})$ plane. 
The details of the calculation are summarized in the caption of Fig.\ \ref{Fig--L_mu_Lambda}. 
In this figure, $(\tilde{\Omega}_{\Lambda}, \tilde{\mu}) = (\Omega_{\Lambda}, 0)$ corresponds to a pure $\Lambda$CDM model for $\Omega_{\Lambda}$.
As illustrated in Fig.\ \ref{Fig--L_mu_Lambda}, the region surrounded by the contours is downward-sloping.
Accordingly,  a low-dissipation model for $\tilde{\Omega}_{\Lambda} = 0.685$ is consistent with a pure $\Lambda$CDM model for a slightly larger value of $\Omega_{\Lambda} (= \tilde{\Omega}_{\Lambda})$.
Of course, the background evolution of the universe is different because it depends on $\tilde{\Omega}_{\Lambda}$. 
In addition, strictly speaking, the low-dissipation model differs from the above pure $\Lambda$CDM model even when density perturbations are examined (see Figs.\ \ref{Fig-delta-a_Lambda} and \ref{Fig-f(z)-z_Lambda}).
However, this result will help to discuss the properties of cosmological models such as an extended $\Lambda$CDM model in a dissipative universe.

In the present study, the radiation-dominated regime has not been discussed because the late universe is focused on.
Interestingly, interacting dark energy models (similar to the modified entropic-force model) suffer from instabilities due to the evolution of matter density perturbations in the radiation-dominated regime (see, e.g., Ref.\ \cite{Roy_2008-2009} and the references therein). 
The instability generally arises from an interaction between dark matter and dark energy, as examined in Ref.\ \cite{Roy_2008-2009}.
Accordingly, the instability probably does not appear in the modified entropic-force model because dark energy is not assumed in this model. 
Of course, the instability should appear in entropic-force models if we assume not only an effective dark energy (discussed in Appendix\ \ref{Modified continuity equation}) but also an interaction between dark matter and effective dark energy. 
Therefore, it is important to examine the entropic-force model in the radiation-dominated regime from different viewpoints. This task is left for the future research.

\section{Conclusions}
\label{Conclusions}

The bulk viscosity of cosmological fluid and the creation of cold dark matter are both able to generate irreversible entropy related to dissipative processes in a homogeneous and isotropic universe.
To examine such a dissipative universe, the general cosmological equations for entropic-force models have been reformulated, focusing on a spatially flat matter-dominated universe. 
Based on this rearranged formulation, the entropic-force term in the acceleration equation can be assumed to include the effects of both reversible and irreversible entropy. 
Using a phenomenological interpretation, a dissipation rate $\tilde{\mu}$ has been defined, and a modified entropic-force model has been developed that includes constant entropic-force terms.
The value of $\tilde{\mu}$ was varied from $0$ to $1$, 
where $\tilde{\mu} =0$ and $\tilde{\mu} =1$ correspond to nondissipative $\Lambda$CDM and fully-dissipative CCDM models, respectively. 
Accordingly, this study bridges the gap between these two standard models.

An effective equation of state parameter $w_{e}$ has been invoked. 
For $\tilde{\mu} =0$, $w_{e}$ is always zero because the effective pressure $p_{e}$ is $0$ in a nondissipative matter-dominated universe. 
However, $w_{e}$ for $\tilde{\mu} >0$ gradually decreases with increasing normalized scale factor $a/a_{0}$ and finally approaches $-1$.
With increasing value of $\tilde{\mu}$, $w_{e}$ decreases.
The dissipation rate $\tilde{\mu}$ affects $w_{e}$ even if the background evolution of the universe is equivalent to that in a fine-tuned standard $\Lambda$CDM model.

Next, the first-order density perturbations in the modified entropic-force model have been analyzed in a neo-Newtonian approach.
The time evolution of the perturbation growth factor $\delta$ has been numerically solved.
When $a/a_{0} \gtrapprox 1$, $\delta$ decreases with increasing $a/a_{0}$, except when $\tilde{\mu} =0$ (which corresponds to a standard $\Lambda$CDM model).
With increasing $\tilde{\mu}$, the perturbation growth factor begins to deviate from the value of $\delta$ for $\tilde{\mu} =0$.
The dissipation rate affects the density perturbations even if the background evolution of the universe remains unchanged. 
However, $\delta$ for $\tilde{\mu}=0.05$ and $0.1$ is not much different from $\delta$ for $\tilde{\mu}=0$ at the present time.

To examine this similarity more closely, the growth rate $f_{c}(z)$ for clustering has been computed.
The calculated values of $f_{c}(z)$ disagree with the observed data points for large dissipation rates  $\tilde{\mu}$, especially at low redshifts.
In particular, $f_{c}(z)$ for $\tilde{\mu} \approx 1$ significantly deviates from the observed data points for low redshift values.
However, $f_{c}(z)$ for low dissipation rates ($\tilde{\mu} \lessapprox 0.1$) agrees with the observations.
Thus, low dissipation rates are suitable for describing structure formations. 
This conclusion has been confirmed by a likelihood analysis for $\tilde{\mu}$.
Therefore, a weakly dissipative universe is a possible scenario. 
The low dissipation rate ($0 < \tilde{\mu} \lessapprox 0.1$) predicts smaller values of $f_{c}(z)$ than does a standard $\Lambda$CDM model (for which $\tilde{\mu} =0$).
Future detailed observations should be able to distinguish a low-dissipation model from a pure $\Lambda$CDM one.

The background evolution of the universe examined so far is the same in every case because $\tilde{\Omega}_{\Lambda}$ is set to be $\Omega_{\Lambda} = 0.685$ in a fine-tuned standard $\Lambda$CDM model.
However, a pure $\Lambda$CDM model for $\Omega_{\Lambda} > 0.685$ may be equivalent to a low-dissipation model for $\tilde{\Omega}_{\Lambda} =0.685$.
Accordingly, density perturbations in the pure $\Lambda$CDM model for $\Omega_{\Lambda} > 0.685$ have been examined as well.
Consequently, when $a/a_{0} \lessapprox 1$, a pure $\Lambda$CDM model for a slightly larger value of $\Omega_{\Lambda}$ is found to be consistent with the low-dissipation model considered here.
However, keep in mind that the low-dissipation model differs from the pure $\Lambda$CDM one in several ways.

The present formulation of a modified entropic-force model is essentially equivalent to an extended $\Lambda$CDM model in a dissipative universe, even though the theoretical backgrounds are different. 
Accordingly, the present model behaves as if a nonzero cosmological constant $\Lambda$ and a dissipative process were operative.
This phenomenological study thereby delineates the properties of cosmological models from different viewpoints.

\appendix

\section{Interpretations of the entropic-force model} 
\label{Modified continuity equation}

As Eqs.\ (\ref{eq:General_FRW01_f_0}) and (\ref{eq:General_FRW02_g_f}) show, the general Friedmann and acceleration equations in a matter-dominated universe (when $p=0$) become
\begin{equation}
   H^2     =  \frac{ 8\pi G }{ 3 } \rho  + f(t)   
\label{eq:General_FRW01_f_2}
\end{equation}
and
\begin{equation}
  \frac{ \ddot{a} }{ a }   
        =  -  \frac{ 4\pi G }{ 3 } \left ( \rho  + \frac{3 p_{e} }{c^2}  \right ) + f(t)   . 
\label{eq:General_FRW02_g_2}
\end{equation}
Substituting Eqs.\ (\ref{eq:General_pe}) and (\ref{eq:Q_0}) into Eq.\ (\ref{eq:drho_General_00}), the general continuity equation becomes
\begin{equation}
       \dot{\rho} + 3  \frac{\dot{a}}{a} \left (  \rho + \frac{p_{e}}{c^2}  \right )   = Q  \rho   ,
\label{eq:drho_General_02_2}
\end{equation}
where $p_{e}$ from Eq.\ (\ref{eq:General_pe}) and $Q$ from Eq.\ (\ref{eq:Q_0}) are 
\begin{equation}
   p_{e} =  - \frac{  c^{2} h(t)} {  4\pi G }    \quad   \textrm{and}   \quad    Q =  - \frac{3}{8 \pi G}  \frac{ \dot{f}(t) }{  \rho  }  .
\label{eq:pe_Q_1}
\end{equation}
This result indicates that Eqs.\ (\ref{eq:General_FRW01_f_2}) to (\ref{eq:pe_Q_1}) are equivalent to those in an extended $\Lambda (t)$CDM model which assumes an effective pressure $p_{e}$ in a dissipative universe.
In other words, the cosmological equations in the present study behave as if a time-varying $\Lambda(t)$ and a dissipative process exist.
When $f(t)$ is a \textrm{constant}, we find that $Q =0$ from Eq.\ (\ref{eq:pe_Q_1}).
Accordingly, the right-hand side of Eq.\ (\ref{eq:drho_General_02_2}) is zero.
The continuity equation is then equivalent to Eq.\ (\ref{eq:drho_General_01_pe}).
In that case, the cosmological equations are equivalent to those of an extended $\Lambda$CDM model in a dissipative universe.

Alternatively, $f(t)$ can be interpreted as effective dark energy.
According to the work of Basilakos and Sol\`{a} \cite{Sola_2014a}, the mass density of the effective dark energy is defined as
\begin{equation}
     \rho_{\textrm{DE}}  \equiv  \frac{3}{8 \pi G} f(t)   .
\label{eq:DE_1}
\end{equation}
In addition, Eqs.\ (\ref{eq:pe_Q_1}) and (\ref{eq:DE_1}) imply 
\begin{equation}
     \dot{\rho}_{\textrm{DE}} = - Q \rho   .
\label{eq:DE_2}
\end{equation}
Substituting Eqs.\ (\ref{eq:DE_1}) and (\ref{eq:DE_2}) into Eqs.\ (\ref{eq:General_FRW01_f_2}) and (\ref{eq:drho_General_02_2}), respectively, and
replacing $\rho$ by $\rho_{m}$, we obtain
\begin{equation}
   H^2     =  \frac{ 8\pi G }{ 3 } ( \rho_{m}  + \rho_{\textrm{DE}}  )   
\label{eq:General_FRW01_f_2_DE}
\end{equation}
and
\begin{equation}
  \dot{\rho}_{\textrm{DE}} +     \dot{\rho}_{m} + 3  \frac{\dot{a}}{a} \left (  \rho_{m} + \frac{p_{e}}{c^2}  \right )   = 0   , 
\label{eq:drho_General_02_2_DE}
\end{equation}
where $\rho_{m}$ is the mass density for matter. 
Equation (\ref{eq:drho_General_02_2_DE}) requires the conservation of the components of the universe, through an exchange of energy between matter and effective dark energy.
A detailed discussion is presented in Ref.\ \cite{Sola_2014a}. 

In this way, the entropic-force model can be usefully interpreted from several fruitful viewpoints.
In the present study, a matter-dominated universe was considered, without assuming any exotic energy component of the universe such as dark energy.
Alternatively, we can assume a phenomenological entropic-force term based on irreversible and reversible entropy terms.


\begin{thebibliography}{99}


\bibitem{Weinberg1} S. Weinberg, \textit{Cosmology} (Oxford University Press, New York, 2008).
\bibitem{Roy1} G. F. R. Ellis, R. Maartens, and  M. A. H. MacCallum, \textit{Relativistic Cosmology} (Cambridge University Press, Cambridge, 2012).
\bibitem{Bamba1_Miao1} K. Bamba, S. Capozziello, S. Nojiri, and S. D. Odintsov, Astrophys. Space Sci. \textbf{342}, 155 (2012); L. Miao, L. Xiao-Dong, W. Shuang, and W. Yi, Commun. Theor. Phys. \textbf{56}, 525 (2011); S. Nojiri and S. D. Odintsov, Int. J. Geom. Methods Mod. Phys. \textbf{04}, 115 (2007).


\bibitem{PERL1998ab} S. Perlmutter \textit{et al.},  Nature (London) \textbf{391}, 51 (1998); Astrophys. J. \textbf{517}, 565 (1999).
\bibitem{Riess1998_2007} A. G. Riess \textit{et al.}, Astron. J. \textbf{116}, 1009 (1998); Astrophys. J. \textbf{607}, 665 (2004); Astrophys. J. \textbf{659},   98 (2007). 
\bibitem{WMAP2011} N. Jarosik \textit{et al.},     Astrophys. J. Suppl. Ser. \textbf{192}, 14 (2011);
                               E. Komatsu \textit{et al.},  Astrophys. J. Suppl. Ser. \textbf{192}, 18 (2011).
\bibitem{Planck2013} P. A. R. Ade \textit{et al.}, arXiv:1303.5076v1 [astro-ph.CO].



\bibitem{Weinberg1989} S. Weinberg, Rev. Mod. Phys. \textbf{61}, 1 (1989);  I. Zlatev, L. Wang, and P. J. Steinhardt, Phys. Rev. Lett.  \textbf{82}, 896 (1999); S. M. Carroll, Living Rev. Relativity \textbf{4}, 1 (2001).



\bibitem{Freese1-Fritzsch1_Overduin1_Sola_2002-2004} K. Freese, F. C. Adams, J. A. Frieman, and E. Mottola, Nucl. Phys. \textbf{B287}, 797 (1987); 
H. A. Borges and S. Carneiro, Gen. Relativ. Gravit. \textbf{37}, 1385 (2005); 
 J. M. Overduin and F. I. Cooperstock, Phys. Rev. D \textbf{58}, 043506 (1998); 
I. L. Shapiro and J. Sol\`{a}, J. High Energy Phys. 02 (2002) 006; 
 I. L. Shapiro, J. Sol\`{a}, C. Espa\~{n}a-Bonet, and P. Ruiz-Lapuente, Phys. Lett. B \textbf{574}, 149 (2003); 
C. Espa\~{n}a-Bonet, P. Ruiz-Lapuente, I. L. Shapiro, and J. Sol\`{a},  J. Cosmol. Astropart. Phys. 02 (2004) 006; 
S. Carneiro, C. Pigozzo, H. A. Borges, and J. S. Alcaniz,  Phys. Rev. D  \textbf{74}, 023532 (2006); 
C. Pigozzo, M. A. Dantas, S. Carneiro, and J. S. Alcaniz,  J. Cosmol. Astropart. Phys. 08 (2011) 022; 
J. S. Alcaniz, H. A. Borges, S. Carneiro, J. C. Fabris, C. Pigozzo, and W. Zimdahl, Phys. Lett. B  \textbf{716}, 165 (2012); 
H. Fritzsch and J. Sol\`{a}, Classical Quantum Gravity \textbf{29}, 215002, (2012); 
J. P. Mimoso and D. Pav\'{o}n, Phys. Rev. D \textbf{87}, 047302 (2013);
J. A. S. Lima, S. Basilakos, and J. Sol\`{a},  Mon. Not. R. Astron. Soc. \textbf{431}, 923 (2013).

\bibitem{Sola_2011a}  J. Grande, J. Sol\`{a}, S. Basilakos, and M. Plionis, J. Cosmol. Astropart. Phys. 08 (2011) 007.
\bibitem{Sola_2009} S. Basilakos, M. Plionis, and J. Sol\`{a}, Phys. Rev. D \textbf{80}, 083511 (2009).
\bibitem{Sola_2011b} J. Sol\`{a},  J. Phys.: Conf. Ser. \textbf{283}, 012033 (2011).
\bibitem{Sola_2013a}  E. L. D. Perico, J. A. S. Lima, S. Basilakos, and J. Sol\`{a}, Phys. Rev. D \textbf{88}, 063531 (2013).
\bibitem{Sola_2013b}  J. Sol\`{a}, J. Phys. Conf. Ser. \textbf{453},  012015 (2013).
\bibitem{Sola_2013c} S. Basilakos and J. Sol\`{a}, Mon. Not. R. Astron. Soc. \textbf{437}, 3331 (2014).
\bibitem{Sola_2014b}  J. Sol\`{a}, AIP Conf. Proc. \textbf{1606}, 19 (2014). 
\bibitem{Lima_2014a} L. L. Graef, F. E. M. Costa, and J. A. S. Lima, Phys. Lett. B \textbf{728}, 400 (2014).
%
\bibitem{Sola_2014c} A. G\'{o}mez-Valent, J. Sol\`{a}, and S. Basilakos, arXiv:1409.7048v2 [astro-ph.CO].






%
\bibitem{Lima_1992e}   J. A. S. Lima and A. S. M. Germano, Phys. Lett. A \textbf{170}, 373 (1992).



\bibitem{Lima-Others1996-2008}   J. A. S. Lima, A. S. M. Germano, and L. R. W. Abramo, Phys. Rev. D \textbf{53}, 4287 (1996).
J.A. S. Lima and J. S. Alcaniz, Astron. Astrophys. \textbf{348}, 1 (1999); 
W. Zimdahl, D. J. Schwarz, A. B. Balakin, and D. Pavon, Phys. Rev. D \textbf{64}, 063501 (2001); 
M. P. Freaza, R. S. de Souza, and I. Waga, Phys. Rev. D \textbf{66}, 103502 (2002); 
J.A. S. Lima, F. E. Silva, and R. C. Santos, Classical Quantum Gravity \textbf{25}, 205006 (2008).

 \bibitem{Lima2010}   J. A. S. Lima, J. F. Jesus, and F. A. Oliveira, J. Cosmol. Astropart. Phys. 11 (2010) 027.
\bibitem{Lima2010b} S. Basilakos, M. Plionis, and J. A. S. Lima, Phys. Rev. D \textbf{82}, 083517 (2010).
\bibitem{Lima2011}  J. F. Jesus, F. A. Oliveira, S. Basilakos, and J. A. S. Lima, Phys. Rev. D \textbf{84}, 063511 (2011).

\bibitem{Lima2012}   J. A. S. Lima, S. Basilakos, and F. E. M. Costa, Phys. Rev. D \textbf{86}, 103534 (2012).
%
\bibitem{Ramos_2014}  R. O. Ramos, M. Vargas dos Santos, and I. Waga, Phys. Rev. D \textbf{89}, 083524 (2014).


\bibitem{Jesus2014}  J. F. Jesus and S. H. Pereira, J. Cosmol. Astropart. Phys. 07 (2014) 040. 

\bibitem{Lima_2014b}   J. A. S. Lima and I. Baranov, Phys. Rev. D \textbf{90}, 043515 (2014).


\bibitem{Prigogine1989} I. Prigogine, J. Geheniau, E. Gunzig, and P. Nardone, Gen. Relativ. Gravit. \textbf{21}, 767 (1989).










%
\bibitem{Weinberg0} S. Weinberg, \textit{Gravitation and Cosmology} (John Wiley \& Sons, New York, 1972). 
\bibitem{Murphy1}   G. L. Murphy, Phys. Rev. D \textbf{8}, 4231 (1973).
\bibitem{Barrow11-Barrow12}  J. D. Barrow, Phys. Lett. B  \textbf{180}, 335 (1986); 
J. D. Barrow, Nucl. Phys. \textbf{B310}, 743 (1988).
%
\bibitem{Davies3}  P. C. W. Davies, Classical Quantum Gravity \textbf{4}, L225 (1987); Ann. Inst. Henri Poincar\'{e}, Sect. A \textbf{49}, 297 (1988).
\bibitem{Lima101} J. A. S. Lima, R. Portugal,  and I. Waga,  Phys. Rev. D \textbf{37}, 2755 (1988).

\bibitem{Zimdahl1-Brevik2} W. Zimdahl, Phys. Rev. D \textbf{53}, 5483 (1996); 
A. I. Arbab, Gen. Relativ. Gravit. \textbf{29}, 61 (1997);
I. Brevik and S. D. Odintsov,  Phys. Rev. D \textbf{65}, 067302 (2002); 
I. Brevik and O. Gorbunova, Gen. Relativ. Gravit. \textbf{37}, 2039 (2005).

\bibitem{Nojiri1-Colistete1} S. Nojiri and S. D. Odintsov, Phys. Rev. D \textbf{72}, 023003 (2005); 
J. Ren and X.-H. Meng, Phys. Lett. B \textbf{633},1 (2006); 
S. Capozziello, V. F. Cardone, E. Elizalde, S. Nojiri, and S. D. Odintsov, Phys. Rev. D \textbf{73}, 043512 (2006); 
J. C. Fabris, S. V. B. Goncalves, and R. de S\'{a} Ribeiro, Gen. Relativ. Gravit. \textbf{38}, 495 (2006); 
R. Colistete, Jr., J. C. Fabris, J. Tossa, and W. Zimdahl, Phys. Rev. D \textbf{76}, 103516 (2007).
\bibitem{Barrow21} B. Li  and  J. D. Barrow, Phys. Rev. D \textbf{79}, 103521 (2009).



\bibitem{Meng2-Avelino1_Odintsov4} X.-H. Meng and X. Dou, Commun. Theor. Phys. \textbf{52}, 377, (2009); arXiv:0812.4904v1 [astro-ph]; 
A. Avelino and U. Nucamendi,  J. Cosmol. Astropart. Phys. 04 (2009) 006; 
W. S. Hipolito-Ricaldi, H. E. S. Velten, and W. Zimdahl,  J. Cosmol. Astropart. Phys. 06 (2009) 016; 
A. Avelino and U. Nucamendi, J. Cosmol. Astropart. Phys. 08 (2010) 009; 
I. Brevik, S. Nojiri, S. D. Odintsov, and D. S\'{a}ez-G\'{o}mez, Eur. Phys. J. C \textbf{69}, 563 (2010); 
O. F. Piattella, J. C. Fabris, and W. Zimdahl, J. Cosmol. Astropart. Phys. 05 (2011) 029; 
X. Dou and X.-H. Meng, Advances in Astronomy \textbf{2011}, 829340 (2011);
O. Pujol\`{a}s, I. Sawicki, and A. Vikman, J. High Energy Phys. 11 (2011) 156;
I. Brevik, E. Elizalde, S. Nojiri, and S. D. Odintsov, Phys. Rev. D \textbf{84}, 103508 (2011).




\bibitem{C_Bulk}
Bulk viscous models assume a bulk viscosity $\xi$ of cosmological fluids and an effective pressure $p_{e}$, which is given by, e.g., $ p_{e} (t) = p(t) - 3 \xi  H(t) $, where $p(t)$ and $H(t)$ represent the pressure of cosmological fluids and the Hubble parameter, respectively.
In general, $\xi$ is assumed to be constant.
However, if $\xi$ is given by $\xi \propto 1/ H$ in a matter-dominated universe (for which $p=0$), we obtain a constant effective pressure, corresponding to a constant driving term for the acceleration equation \cite{Koma5,Koma6}.



\bibitem{C_Bulk_CCDM}
An equivalence of the bulk viscosity and matter creation dissipative mechanisms has been discussed in Ref.\ \cite{Lima_1992e}.



















\bibitem{Easson1}  D. A. Easson, P. H. Frampton, and G. F. Smoot, Phys. Lett. B \textbf{696}, 273 (2011).
\bibitem{Easson2}  D. A. Easson, P. H. Frampton, and G. F. Smoot, Int. J. Mod. Phys. A \textbf{27}, 1250066 (2012).


\bibitem{Koivisto1_Cai1_Cai2_Qiu1_Casadio1-Costa1}  
T. S. Koivisto, D. F. Mota, and M. Zumalac\'{a}rregui, J. Cosmol. Astropart. Phys. 02 (2011) 027; 
Y. F. Cai, J. Liu, and H. Li,      Phys. Lett. B \textbf{690}, 213 (2010);  
Y. F. Cai and E. N. Saridakis, Phys. Lett. B \textbf{697}, 280 (2011); 
T. Qiu and E. N. Saridakis,     Phys. Rev. D \textbf{85}, 043504 (2012); 
%
R. Casadio and A. Gruppuso, Phys. Rev. D \textbf{84}, 023503 (2011); 
U. H. Danielsson, arXiv:1003.0668v1 [hep-th]; 
Y. S. Myung, Astrophys. Space Sci. \textbf{335}, 553 (2011); 
F. E. M. Costa, J. A. S. Lima, and F. A. Oliveira, arXiv:1204.1864v1 [astro-ph.CO].

\bibitem{Basilakos1}
 S. Basilakos, D. Polarski, and J. Sol\`{a}, Phys. Rev. D \textbf{86}, 043010 (2012).

\bibitem{Lepe1} S. Lepe and F. Pen\~{a}, arXiv:1201.5343v2 [hep-th].

\bibitem{Koma4}  N. Komatsu and S. Kimura, Phys. Rev. D \textbf{87}, 043531 (2013); N. Komatsu, JPS Conf. Proc. \textbf{1}, 013112 (2014).
\bibitem{Koma5}  N. Komatsu and S. Kimura, Phys. Rev. D \textbf{88}, 083534 (2013).
\bibitem{Koma6}  N. Komatsu and S. Kimura, Phys. Rev. D \textbf{89}, 123501 (2014).

\bibitem{Sola_2014a} S. Basilakos and J. Sol\`{a}, Phys. Rev. D \textbf{90}, 023008 (2014).




\bibitem{Prigogine_1998} D. Kondepudi and I. Prigogine, \textit{Modern Thermodynamics: From Heat Engines to Dissipative Structures} (John Wiley \& Sons, New York, 1998).




\bibitem{Padma1}  T. Padmanabhan, Mod. Phys. Lett. A \textbf{25}, 1129 (2010).
\bibitem{Verlinde1} E. Verlinde, J. High Energy Phys. 04 (2011) 029.




\bibitem{Barrow22}  J. D. Barrow and T. Clifton, Phys. Rev. D \textbf{73}, 103520 (2006). 

\bibitem{Amendola1Zimdahl01}  L. Amendola, Phys. Rev. D \textbf{62}, 043511 (2000); 
W. Zimdahl, D. Pav\'{o}n, and L. P. Chimento, Phys. Lett. B  \textbf{521}, 133 (2001).



\bibitem{Wang0102_YWang2014}  B. Wang, Y. Gong, and E. Abdalla, Phys. Lett. B  \textbf{624}, 141 (2005); 
B. Wang, C.-Y. Lin, D. Pav\'{o}n, and E. Abdalla, Phys. Lett. B  \textbf{662}, 1 (2008); 
Y. Wang, D. Wands, G-B. Zhao, and L. Xu, Phys. Rev. D \textbf{90}, 023502 (2014).


\bibitem{Pavon_2005} D. Pav\'{o}n and W. Zimdahl, Phys. Lett. B \textbf{628}, 206 (2005);
M. R. Setare, Phys. Lett. B \textbf{642}, 1 (2006); 
B. Hu and Y. Ling, Phys. Rev. D \textbf{73}, 123510 (2006).









\bibitem{Bekenstein1}  J. D. Bekenstein, Phys. Rev. D \textbf{7}, 2333 (1973); Phys. Rev. D \textbf{9}, 3292 (1974);  Phys. Rev. D \textbf{12}, 3077 (1975).

\bibitem{Tsallis2012}  C. Tsallis and L. J. L. Cirto, Eur. Phys. J. C \textbf{73}, 2487 (2013).
\bibitem{Tsa0}    C. Tsallis, J. Stat. Phys. \textbf{52}, 479 (1988).
\bibitem{Tsa1}    C. Tsallis,  {\it Introduction to Nonextensive Statistical Mechanics: Approaching a Complex World} (Springer, New York, 2009).








\bibitem{Hooft-Bousso}
G. 't Hooft, arXiv:gr-qc/9310026; L. Susskind, J. Math. Phys. \textbf{36}, 6377 (1995); R. Bousso, Rev. Mod. Phys. \textbf{74}, 825 (2002).







\bibitem{Lima_Newtonian_1997} J. A. S. Lima, V. Zanchin, and R. Brandenberger, Mon. Not. R. Astron. Soc. \textbf{291}, L1 (1997).

\bibitem{McCrea_1951} W. H. McCrea, Proc. R. Soc. A \textbf{206}, 562 (1951).
\bibitem{Harrison_1965} E. R. Harrison, Ann. Phys. (N.Y.) \textbf{35}, 437 (1965).





\bibitem{C4} 
In a $\Lambda (t)$CDM model \cite{Sola_2009}, the time dependence of $\Lambda(t)$ appears always at the expense of an interaction with matter. 
This model can be interpreted as an energy exchange cosmology which assumes the transfer of energy between two fluids.







\bibitem{Waga1994}  R. C. Arcuri and I. Waga, Phys. Rev. D \textbf{50}, 2928 (1994).

\bibitem{Sola_2007-2009}
J. Grande, R. Opher, A. Pelinson, and J. Sol\`{a}, J. Cosmol. Astropart. Phys. 12 (2007) 007; 
J. Grande, A. Pelinson, and J. Sol\`{a}, Phys. Rev. D \textbf{79}, 043006 (2009).  











\bibitem{Reis_2003}   R. R. R. Reis, Phys. Rev. D \textbf{67}, 087301 (2003); \textbf{68}, 089901(E) (2003).





\bibitem{Odintsov_2006_b}  S. Nojiri and S. D. Odintsov, Phys. Lett. B \textbf{639}, 144 (2006).





%
\bibitem{Peebles_1993} P. J. E. Peebles, \textit{Principles of Physical Cosmology} (Princeton University Press, Princeton, 1993).





\bibitem{Colless2001} M. Colless \textit{et al.}, Mon. Not. R. Astron. Soc. \textbf{328}, 1039 (2001).
\bibitem{Guzzo2008} L. Guzzo \textit{et al.}, Nature (London) \textbf{451}, 541 (2008).
\bibitem{Tegmark2006} M. Tegmark \textit{et al.}, Phys. Rev. D \textbf{74}, 123507 (2006).
\bibitem{Ross2007} N. P. Ross \textit{et al.}, Mon. Not. R. Astron. Soc. \textbf{381}, 573 (2007).
\bibitem{Angela2007} J. da Angela \textit{et al.}, Mon. Not. R. Astron. Soc. \textbf{383}, 565 (2008).
\bibitem{Viel2004} M. de Viel, M. G. Haehnelt, and V. Springel, Mon. Not. R. Astron. Soc. \textbf{354}, 684 (2004).
\bibitem{McDonald2005} P. McDonald \textit{et al.}, Astrophys. J. \textbf{635}, 761 (2005).






\bibitem{Roy_2008-2009}
J. V\"{a}liviita, E. Majerotto, and R. Maartens, J. Cosmol. Astropart. Phys. 07 (2008) 020; 
E. Majerotto, J. V\"{a}liviita, and R. Maartens, Nucl. Phys. B, Proc. Suppl. \textbf{194}, 260 (2009).

















\end{thebibliography}
\end{document}